\documentclass[varenna,final]{cimento}
  
\usepackage{amsmath,amssymb}
\usepackage{graphicx}

\graphicspath{{.}{./figures/}}

\begin{document}

\title{Ab initio computations of atomic nuclei}
\shorttitle{Ab initio computations of atomic nuclei}

\author{T.~Papenbrock}

\institute{Department of Physics and Astronomy, University of Tennessee, Knoxville, TN 37996, USA \\ 
Physics Division, Oak Ridge National Laboratory, Oak Ridge, TN 37831, USA}

\maketitle

\begin{abstract}
Ab initio computations of atomic nuclei, based on Hamiltonians from effective field theories of quantum chromodynamics, are now routinely used to predict and describe properties of medium heavy nuclei, and even the heavy nucleus $^{208}$Pb has been reached. These lecture notes describe what are the central ideas and concepts behind the Hamiltonians and some of the methods that have enabled this progress. 
\end{abstract}


\newpage
\tableofcontents
\newpage

\section{Introduction}
\label{sec:intro}
In the past three decades, ab initio computations have moved from applications in nuclear few-body systems to medium-mass and even heavy nuclei. An increasing number of researchers is using and advancing such computations and their popularity is also due to their affordability.  Many different methods are employed in this endeavor, ranging from virtually exact methods such as Green's function Monte Carlo~\cite{carlson2015} and the no-core shell model~\cite{navratil2009,barrett2013} to powerful approximations such as coupled-cluster theory~\cite{hagen2014}, Gorkov-Green's function methods~\cite{dickhoff2004,soma2013}, the in-medium similarity renormalization group (IMSRG)~\cite{hergert2016}, and nuclear lattice effective field theory~\cite{lee2009}. The effort for the exact methods increases exponentially with increasing mass number  and -- meeting exponentially increasing computer power due to Moore's law -- have approximately advanced linearly in mass number as time has progressed.  In contrast, the approximate methods exhibit a cost that increases polynomially with increasing system size. They have moved the frontier of ab initio computations from light nuclei towards increasingly heavier nuclei~\cite{morris2018,arthuis2020} and, recently, up to $^{208}$Pb~\cite{hu2022}. These lecture notes  focus on them. 

A main argument is that computing nuclei from Hamiltonians is not exponentially hard, because the low-energy structure of many nuclei is simple enough that it can be computed sufficiently accurately and in a controlled fashion using methods that scale affordably. Examples are energies, radii, and electromagnetic moments of ground states and low-energy excitations. Of course, there are certainly observables, or nuclei, for which computations do not scale favorably. In general it is hard to compute finely tuned or ``small'' quantities accurately, or nuclei whose structure is not simple. Compound-nucleus states, for instance, are thought to have a very complex structure and their statistics is indistinguishable from those of random matrices~\cite{mitchell2010}. It would seem hard to compute those accurately.  
While the simple structure and the regular features of many nuclei can be computed efficiently (i.e., without requiring resources that scale exponentially with increasing mass number) such computations are not inexpensive and often require leadership-class computing resources. Nevertheless, they are affordable and can be done.   

Another argument in favor of approximate solutions to the nuclear many-body problem is that exact computations are rarely required. Nuclear Hamiltonians are only approximate themselves. More precisely: they are effective Hamiltonians for approximate descriptions of low-energy nuclear phenomena. When taken from effective theories of quantum chromodynamics~\cite{bedaque2002,epelbaum2009,hammer2020}, effective Hamiltonians are worked out up to some order in the power counting. Then it is not necessary to solve such  approximate Hamiltonians exactly. All one really needs is that the error in the solution is smaller than the uncertainty of the Hamiltonian for the observables of interest.    

As the reader will see, ab initio computations are heuristic approaches that build on phenomenology and the insights gained from simple models. On first sight, this might be not what one would expect from ab initio methods.  However, heuristics enter almost everywhere in heavy nuclei because the states one is interested in are -- strictly speaking -- metastable. The ground-state of the nucleus $^{208}$Pb, for instance, is not the state of lowest energy for a system of 82 protons and 126 neutrons, because it has a positive $\alpha$-particle separation energy (as has the majority of nuclei above tin). Indeed, only nuclei below proton number $Z\approx 40$ are bound with respect to decay into cluster. Thus most nuclei are only metastable resonances, held together by enormous Coulomb barriers. Deformed nuclei are another example. Here, phenomenology taught us that quadrupole (and possibly octopole) deformations are most important. It would be hard to significantly enlarge the set of constraints that are probed in mean-field calculations to find the most appropriate product state of a nucleus. In other words, the search for the optimal reference state itself might already be exponentially hard if one were not to take known phenomenology into account. Thus, practitioners of ab initio computations very much benefit from the collected wisdom of nuclear structure and many-body methods~\cite{bohr1975,ringschuck}.

The ideas and concepts behind some of today's ab initio computations date back to the 1960s.  Coester and K\"ummel~\cite{coester1960}, for instance, introduced coupled-cluster theory as a method to compute ``short-range correlations in nuclear wave
functions.''  These correlations deliver the bulk of nuclear binding energy. Lipkin~\cite{lipkin1960} pointed out the connection between collective excitations and symmetry breaking. However, it took decades to computationally realize this vision and to base such calculations on interactions that have a microscopic foundation in quantum chromodynamics~\cite{epelbaum2009,hammer2020}.

These lecture notes are organized as follows. They start with a definition of the term ``ab initio'' in Sect.~\ref{sec:abinitio}. Next follows a brief introduction of Hamiltonians from effective field theories in Sect.~\ref{sec:hams}. Necessary details about single-particle bases (e.g. the number of single-particle states required for a computation and the computation of matrix elements) are presented in Sect.~\ref{sec:sp}. Section~\ref{sec:hf} is dedicated to the importance of the mean-field and  symmetry breaking. The inclusion of correlations is presented in Sect.~\ref{sec:ccm}, and excited states are discussed in Sect.~\ref{sec:excited}. Finally a few results are highlighted in Sect.~\ref{sec:impact} and the Epilogue~\ref{sec:summary} concludes these lecture notes.

\section{What does ``ab initio'' mean?}
\label{sec:abinitio}
Quantum chromodynamics is the theory of the strong nuclear force, and this then literally is the beginning from which nuclear theory could start. However, there are practical and principal objections to this starting point. At present, computations of lattice quantum chromodynamics (at zero temperature) are limited to one or two hadrons, and computations of all but the lightest nuclei seem impractical in the foreseeable future. Principal objections against the reductionist approach~\cite{anderson1972} concern the emergence of physical phenomena, e.g., the spontaneous breaking of chiral symmetry in quantum chromodynamics or the occurrence of deformation and superfluidity in finite nuclei, that are hard or impossible to see in simulations of finite systems. 

\begin{figure}
\includegraphics[width=0.5\textwidth]{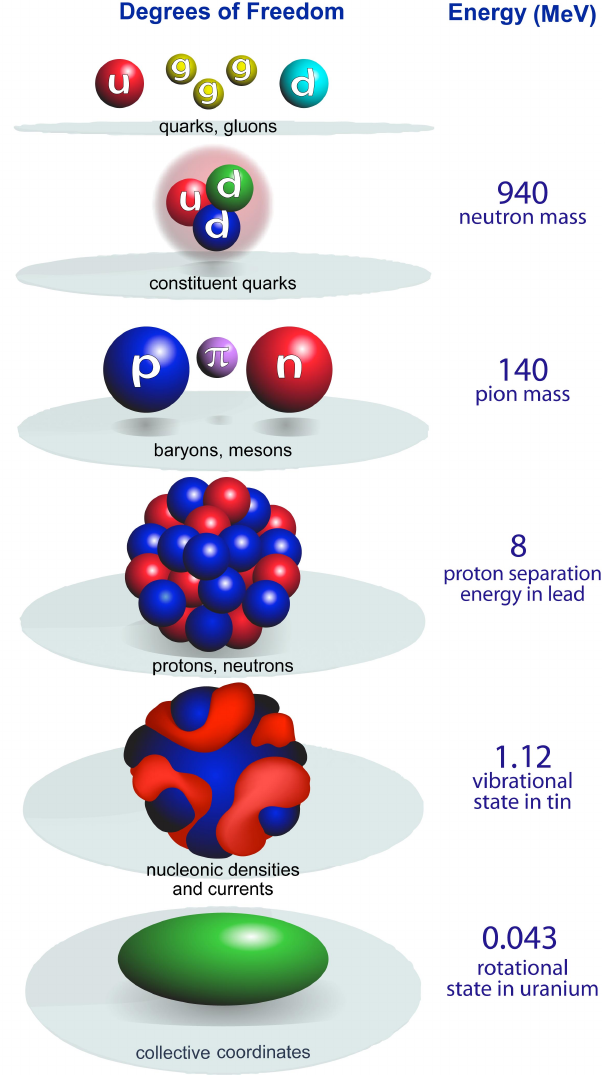}
\caption{Energy scales and corresponding effective degrees of freedom in nuclear physics. Adapted from Ref.~\cite{bertsch2007} with permission from Witek Nazarewicz, see also Ref.~\cite{nazarewicz2016}.}
\label{fig:scales}
\end{figure}

Figure~\ref{fig:scales} shows energy scales and corresponding effective degrees of freedom in nuclear physics. At highest energies in excess of 1~GeV, one deals with quantum chromodynamics and  quarks and gluons are the relevant degrees. At lower energies, quarks are confined into hadrons and  chiral symmetry is spontaneously broken; the relevant degrees of freedom are nucleons (protons and neutrons) and pions. This is where chiral effective field theory comes in. Nuclei also exhibit phenomena at  different energy scales. It takes about 8~MeV to separate a nucleon from a nucleus (and the total binding energy is about 8~MeV per nucleon). Low-energy nuclear excitations are collective vibrations and -- at lowest energies -- rotations of the whole nucleus. Densities and Euler angles are the relevant degrees of freedom to describe vibrations and rotations, respectively. Figure~\ref{fig:scales} also depicts that the resolution decreases with decreasing energy.  

The separation of energy scales depicted in Fig.~\ref{fig:scales} suggests that there are various opportunities for effective theories or models that capture the phenomena at a given resolution scale. Where to start then?  
To address this question, one needs to balance resolution scale and predictive power. Ideally, one would want both to be as large as possible.  
The highest resolution would be achieved when starting at the top, i.e. by using quantum chromodynamics, the fundamental theory of the strong nuclear force that acts between quarks and gluons. Lattice quantum chromodynamics, i.e. a numerical approach that discretizes space and time and uses Monte Carlo techniques to compute observable quantities, has made impressive progress over the past two decades. Examples include the computation of hadron spectra~\cite{durr2008}, and -- when combined with quantum electrodynamics -- the proton-neutron mass difference~\cite{borsanyi2014}. However, computing bound states between two baryons is a much more challenging task. Many of such calculations use unphysical (larger) pion masses to facilitate the computation, and it was controversial whether nuclear binding decreases~\cite{aoki2010,aoki2012} or increases~\cite{beane2012,beane2013} with increasing pion mass, see Ref.~\cite{drischler2021} for a discussion. However, recent progress~\cite{green2021,nicholson2022} seems to indicate that a resolution of the controversy is coming soon.
In any case, nuclear computations at the highest resolution are only reliable for one or two hadrons, and all but the lightest nuclei seem to be out of reach of lattice quantum chromodynamics for some time. Thus, at present this high-resolution approach lacks predictive power for most nuclei. 
Let me contrast this with computations based on nucleonic densities and currents~\cite{vretanar2005,delaroche2010,niksic2011,erler2012}. These start at a lower resolution (and the connection to quantum chromodynamics is obscured) but have predictive power for nuclei  across the nuclear chart.  

This discussion made clear that there is a tension between high resolution and high predictive power. One can use this to define the expression ``ab initio.''  This lecture follows the authors of Ref.~\cite{ekstrom2023} who ``interpret(ed) the ab
initio method to be a systematically improvable approach for quantitatively describing nuclei
using the finest resolution scale possible while maximizing its predictive capabilities.'' Today such a program starts with chiral effective field theory to ground the interactions between two and three nucleons to quantum chromodynamics and then uses systematically improvable approximations to solve the nuclear quantum many-body problem.

\section{Hamiltonians from effective field theories of QCD}
\label{sec:hams}
Chiral effective field theory builds on ideas by Weinberg~\cite{weinberg1990,weinberg1991} that were put into action by many researchers, see Refs.~\cite{ordonez1992,park1993,ordonez1994,vankolck1994,ordonez1996,park1996,kaiser1997,kaiser1998,bernard1997,epelbaoum1998,epelbaum2000,epelbaum2002,entem2003}. The approach is reviewed in Refs.~\cite{epelbaum2009,machleidt2011,hammer2020}. Here, I only present the main ideas and briefly summarize successes and challenges. More details can be found, e.g., in the textbook~\cite{scherer2012}.

\subsection{Spontaneous breaking of chiral symmetry}
It is useful to consider quantum chromodynamics with only the two lightest quarks, $u$ and $d$. Quarks come in four-spinors (that are also denoted simply by $u$ and $d$) consisting of pairs of a left-handed and a right-handed two-spinor, i.e. one has
\begin{equation}
\begin{aligned}
    u &= u_L + u_R \ , \\
    d &= d_L + d_R \ .
\end{aligned}
\end{equation}
Here, left- and right-handed refers to chirality, i.e. to different transformation properties under Lorentz transformations. For massless quarks, this chirality is identical to the familiar helicity, i.e. the fixed projection of the massless fermion's spin onto its momentum. If one neglects the small masses of the $u$ and $d$ quarks, the QCD Lagrangian becomes invariant under independent SU$(2)$ rotations between (chiral) left and right-handed quarks, i.e. one can rotate left-handed $u$ and $d$ quarks using one SU$(2)_L$ transformation and rotate right-handed $u$ and $d$ quarks using another SU$(2)_R$ transformation without changing the Lagrangian. This is the chiral symmetry exhibited by the massless QCD Lagrangian.

Instead of using transformations SU$(2)_L$ and SU$(2)_R$, one can also consider two transformations SU$(2)_{L+R}$ and SU$(2)_{L-R}$ where one either rotates $u$ and $d$ quarks together or against each other. (The former are known as ``vector'' and the latter as ``axial-vector'' transformations.) The SU$(2)_{L+R}$ symmetry operation has isospin as its conserved quantity, and its impact on the hadron spectrum is well known: The nucleon is an isospin-1/2 doublet and the $\Delta$ resonance is an isospin-3/2 quartet of states. (Note that isospin symmetry is also exact for finite but equal masses of $u$ and $d$ quarks.) However, the hadron spectrum exhibits no other degeneracies that could be identified with the axial-vector SU$(2)_{L-R}$ symmetry. Thus, it is thought that the QCD vacuum spontaneously breaks the axial-vector symmetry SU$(2)_{L-R}$. This is the spontaneous breaking of chiral symmetry in QCD.

Spontaneous symmetry breaking has impactful consequences: Nambu-Goldstone bosons are the low-lying excitations in such systems. As SU$(2)_{L-R}$ has three generators, one identifies the three charge states of the pion as the Nambu-Goldstone bosons of the spontaneously broken chiral symmetry. The fact that pions are not massless is understood because the masses of the $u$ and $d$ quarks are only small, but not zero. Thus, there is spontaneous and explicit breaking of the axial SU$(2)_{L-R}$ symmetry.   

Some authors consider the $u$, $d$, and $s$ quarks to be light~\cite{scherer2012}. Then the arguments made above can be extended to this case and one essentially replaces SU$(2)$ by SU$(3)$. That group has eight generators and the Nambu Goldstone bosons comprise -- in addition to the three pions --  four kaon states and the eta meson. The kaons have masses of about 500~MeV and the eta of about 550~MeV. These masses are already about half the nucleon mass, and they are not treated explicitly in effective theories of the nuclear force.   

From a theorist's perspective, spontaneous symmetry breaking is also great news because Nambu-Goldstone bosons are weakly interacting and the effective field theories that describe their interactions (and those with other hadrons) are severely constrained. This has been worked out by Weinberg~\cite{weinberg1968} for the case of pions and led to chiral perturbation theory, see the textbook~\cite{scherer2012} for details. Weinberg's idea for nuclear physics~\cite{weinberg1990,weinberg1991} was then to construct effective nuclear interactions using pion exchange (as in chiral perturbation theory) for the long-range part of the nuclear force and short-range ``contact'' potentials for the unknown short-range part of the interaction. The Weinberg power counting is for the potential. It is in powers ${\rm max}(q,m_\pi)/{\rm min}(\Lambda,\Lambda_b)$ where $q$ is the momentum transfer probed, $m_\pi$ the pion mass, $\Lambda_b\sim{\cal O}(1~{\rm GeV})$ the breakdown scale, and $\Lambda$ the employed momentum cutoff.   

\subsection{How effective field theories work} One can understand how and why the description of unknown short-range physics in terms of contact potentials works. The following exposition is based on the lectures by Lepage~\cite{lepage1997}. He considered a problem where the potential $V(r)=V_{\rm long}(r) +V_{\rm short}(r)$ consisted of a known long-range part (which was taken to be the Coulomb potential $V_{\rm long}(r)=-\alpha/r$ and an undisclosed short range part $V_{\rm short}$. However, ``data'' consisting of scattering phase shifts of the potentials $V$ at low energies were disclosed. In an effective field theory one can then use the effective potential 
\begin{equation}
\label{lepage}
    V_{\rm eff}=V_{\rm long} + c_1 a^2\delta_a(\mathbf{r}) - c_2 a^4\nabla^2\delta_a(\mathbf{r}) +\ldots .
\end{equation}
Here, $a$ is a short range and $\delta_a(\mathbf{r})$ is a smeared $\delta$-function (with range $a$); thus $1/a$ serves as a momentum cutoff. The low-energy constants $c_1$ and $c_2$ are the strength of the leading-order and next-to-leading-order terms, respectively and they have to be adjusted to data. The specific powers of $a$ are chosen such that low-energy constants are dimensionless.  When acting onto plane waves with wave number $k$, the next-to-leading order term is of size  ${\cal O}((ak)^2)$ compared to the leading-order one. Thus, for a sufficiently low momentum $k$ one has $ak\ll 1$ and the contribution from next-to-leading order is much smaller than that from the leading order.  This makes clear how contact terms (and derivatives acting upon them) provide us with a systematically improvable approach to long wavelength phenomena.    

\begin{figure}
\begin{center}
\setlength{\unitlength}{0.240900pt}
\ifx\plotpoint\undefined\newsavebox{\plotpoint}\fi
\sbox{\plotpoint}{\rule[-0.200pt]{0.400pt}{0.400pt}}%
\begin{picture}(1349,749)(0,0)
\font\gnuplot=cmr10 at 10pt
\gnuplot
\sbox{\plotpoint}{\rule[-0.200pt]{0.400pt}{0.400pt}}%
\put(220.0,249.0){\rule[-0.200pt]{4.818pt}{0.400pt}}
\put(198,249){\makebox(0,0)[r]{$10^{-6}$}}
\put(1265.0,249.0){\rule[-0.200pt]{4.818pt}{0.400pt}}
\put(220.0,385.0){\rule[-0.200pt]{4.818pt}{0.400pt}}
\put(198,385){\makebox(0,0)[r]{$10^{-4}$}}
\put(1265.0,385.0){\rule[-0.200pt]{4.818pt}{0.400pt}}
\put(220.0,522.0){\rule[-0.200pt]{4.818pt}{0.400pt}}
\put(198,522){\makebox(0,0)[r]{$10^{-2}$}}
\put(1265.0,522.0){\rule[-0.200pt]{4.818pt}{0.400pt}}
\put(220.0,658.0){\rule[-0.200pt]{4.818pt}{0.400pt}}
\put(198,658){\makebox(0,0)[r]{$1$}}
\put(1265.0,658.0){\rule[-0.200pt]{4.818pt}{0.400pt}}
\put(220.0,113.0){\rule[-0.200pt]{0.400pt}{2.409pt}}
\put(220.0,716.0){\rule[-0.200pt]{0.400pt}{2.409pt}}
\put(243.0,113.0){\rule[-0.200pt]{0.400pt}{2.409pt}}
\put(243.0,716.0){\rule[-0.200pt]{0.400pt}{2.409pt}}
\put(263.0,113.0){\rule[-0.200pt]{0.400pt}{2.409pt}}
\put(263.0,716.0){\rule[-0.200pt]{0.400pt}{2.409pt}}
\put(280.0,113.0){\rule[-0.200pt]{0.400pt}{2.409pt}}
\put(280.0,716.0){\rule[-0.200pt]{0.400pt}{2.409pt}}
\put(295.0,113.0){\rule[-0.200pt]{0.400pt}{2.409pt}}
\put(295.0,716.0){\rule[-0.200pt]{0.400pt}{2.409pt}}
\put(309.0,113.0){\rule[-0.200pt]{0.400pt}{4.818pt}}
\put(309,68){\makebox(0,0){0.001}}
\put(309.0,706.0){\rule[-0.200pt]{0.400pt}{4.818pt}}
\put(398.0,113.0){\rule[-0.200pt]{0.400pt}{2.409pt}}
\put(398.0,716.0){\rule[-0.200pt]{0.400pt}{2.409pt}}
\put(450.0,113.0){\rule[-0.200pt]{0.400pt}{2.409pt}}
\put(450.0,716.0){\rule[-0.200pt]{0.400pt}{2.409pt}}
\put(487.0,113.0){\rule[-0.200pt]{0.400pt}{2.409pt}}
\put(487.0,716.0){\rule[-0.200pt]{0.400pt}{2.409pt}}
\put(516.0,113.0){\rule[-0.200pt]{0.400pt}{2.409pt}}
\put(516.0,716.0){\rule[-0.200pt]{0.400pt}{2.409pt}}
\put(539.0,113.0){\rule[-0.200pt]{0.400pt}{2.409pt}}
\put(539.0,716.0){\rule[-0.200pt]{0.400pt}{2.409pt}}
\put(559.0,113.0){\rule[-0.200pt]{0.400pt}{2.409pt}}
\put(559.0,716.0){\rule[-0.200pt]{0.400pt}{2.409pt}}
\put(576.0,113.0){\rule[-0.200pt]{0.400pt}{2.409pt}}
\put(576.0,716.0){\rule[-0.200pt]{0.400pt}{2.409pt}}
\put(591.0,113.0){\rule[-0.200pt]{0.400pt}{2.409pt}}
\put(591.0,716.0){\rule[-0.200pt]{0.400pt}{2.409pt}}
\put(605.0,113.0){\rule[-0.200pt]{0.400pt}{4.818pt}}
\put(605,68){\makebox(0,0){0.01}}
\put(605.0,706.0){\rule[-0.200pt]{0.400pt}{4.818pt}}
\put(694.0,113.0){\rule[-0.200pt]{0.400pt}{2.409pt}}
\put(694.0,716.0){\rule[-0.200pt]{0.400pt}{2.409pt}}
\put(746.0,113.0){\rule[-0.200pt]{0.400pt}{2.409pt}}
\put(746.0,716.0){\rule[-0.200pt]{0.400pt}{2.409pt}}
\put(783.0,113.0){\rule[-0.200pt]{0.400pt}{2.409pt}}
\put(783.0,716.0){\rule[-0.200pt]{0.400pt}{2.409pt}}
\put(811.0,113.0){\rule[-0.200pt]{0.400pt}{2.409pt}}
\put(811.0,716.0){\rule[-0.200pt]{0.400pt}{2.409pt}}
\put(835.0,113.0){\rule[-0.200pt]{0.400pt}{2.409pt}}
\put(835.0,716.0){\rule[-0.200pt]{0.400pt}{2.409pt}}
\put(855.0,113.0){\rule[-0.200pt]{0.400pt}{2.409pt}}
\put(855.0,716.0){\rule[-0.200pt]{0.400pt}{2.409pt}}
\put(872.0,113.0){\rule[-0.200pt]{0.400pt}{2.409pt}}
\put(872.0,716.0){\rule[-0.200pt]{0.400pt}{2.409pt}}
\put(887.0,113.0){\rule[-0.200pt]{0.400pt}{2.409pt}}
\put(887.0,716.0){\rule[-0.200pt]{0.400pt}{2.409pt}}
\put(900.0,113.0){\rule[-0.200pt]{0.400pt}{4.818pt}}
\put(900,68){\makebox(0,0){0.1}}
\put(900.0,706.0){\rule[-0.200pt]{0.400pt}{4.818pt}}
\put(989.0,113.0){\rule[-0.200pt]{0.400pt}{2.409pt}}
\put(989.0,716.0){\rule[-0.200pt]{0.400pt}{2.409pt}}
\put(1041.0,113.0){\rule[-0.200pt]{0.400pt}{2.409pt}}
\put(1041.0,716.0){\rule[-0.200pt]{0.400pt}{2.409pt}}
\put(1078.0,113.0){\rule[-0.200pt]{0.400pt}{2.409pt}}
\put(1078.0,716.0){\rule[-0.200pt]{0.400pt}{2.409pt}}
\put(1107.0,113.0){\rule[-0.200pt]{0.400pt}{2.409pt}}
\put(1107.0,716.0){\rule[-0.200pt]{0.400pt}{2.409pt}}
\put(1130.0,113.0){\rule[-0.200pt]{0.400pt}{2.409pt}}
\put(1130.0,716.0){\rule[-0.200pt]{0.400pt}{2.409pt}}
\put(1150.0,113.0){\rule[-0.200pt]{0.400pt}{2.409pt}}
\put(1150.0,716.0){\rule[-0.200pt]{0.400pt}{2.409pt}}
\put(1167.0,113.0){\rule[-0.200pt]{0.400pt}{2.409pt}}
\put(1167.0,716.0){\rule[-0.200pt]{0.400pt}{2.409pt}}
\put(1182.0,113.0){\rule[-0.200pt]{0.400pt}{2.409pt}}
\put(1182.0,716.0){\rule[-0.200pt]{0.400pt}{2.409pt}}
\put(1196.0,113.0){\rule[-0.200pt]{0.400pt}{4.818pt}}
\put(1196,68){\makebox(0,0){1}}
\put(1196.0,706.0){\rule[-0.200pt]{0.400pt}{4.818pt}}
\put(1285.0,113.0){\rule[-0.200pt]{0.400pt}{2.409pt}}
\put(1285.0,716.0){\rule[-0.200pt]{0.400pt}{2.409pt}}
\put(220.0,113.0){\rule[-0.200pt]{256.558pt}{0.400pt}}
\put(1285.0,113.0){\rule[-0.200pt]{0.400pt}{147.672pt}}
\put(220.0,726.0){\rule[-0.200pt]{256.558pt}{0.400pt}}
\put(45,464){\makebox(0,0){$|\Delta \delta_0(E)|$}}
\put(752,23){\makebox(0,0){$E$}}
\put(361,658){\makebox(0,0)[l]{$-\alpha/r$}}
\put(361,501){\makebox(0,0)[l]{$-\alpha/r+c_1a^2\delta_a(\mathbf{r})$}}
\put(605,297){\makebox(0,0)[l]{$-\alpha/r+c_1a^2\delta_a(\mathbf{r})-c_2a^4\nabla^2\delta_a(\mathbf{r})$}}
\put(220.0,113.0){\rule[-0.200pt]{0.400pt}{147.672pt}}
\put(309,323){\usebox{\plotpoint}}
\multiput(309.00,323.58)(1.416,0.498){97}{\rule{1.228pt}{0.120pt}}
\multiput(309.00,322.17)(138.451,50.000){2}{\rule{0.614pt}{0.400pt}}
\multiput(450.00,373.58)(1.303,0.498){81}{\rule{1.138pt}{0.120pt}}
\multiput(450.00,372.17)(106.638,42.000){2}{\rule{0.569pt}{0.400pt}}
\multiput(559.00,415.58)(1.104,0.496){39}{\rule{0.976pt}{0.119pt}}
\multiput(559.00,414.17)(43.974,21.000){2}{\rule{0.488pt}{0.400pt}}
\multiput(605.00,436.58)(1.540,0.498){89}{\rule{1.326pt}{0.120pt}}
\multiput(605.00,435.17)(138.248,46.000){2}{\rule{0.663pt}{0.400pt}}
\multiput(746.00,482.58)(1.964,0.497){53}{\rule{1.657pt}{0.120pt}}
\multiput(746.00,481.17)(105.561,28.000){2}{\rule{0.829pt}{0.400pt}}
\multiput(855.00,510.59)(4.940,0.477){7}{\rule{3.700pt}{0.115pt}}
\multiput(855.00,509.17)(37.320,5.000){2}{\rule{1.850pt}{0.400pt}}
\multiput(900.00,515.58)(1.729,0.498){79}{\rule{1.476pt}{0.120pt}}
\multiput(900.00,514.17)(137.937,41.000){2}{\rule{0.738pt}{0.400pt}}
\multiput(1041.00,556.58)(2.203,0.497){47}{\rule{1.844pt}{0.120pt}}
\multiput(1041.00,555.17)(105.173,25.000){2}{\rule{0.922pt}{0.400pt}}
\multiput(1150.00,581.58)(2.372,0.491){17}{\rule{1.940pt}{0.118pt}}
\multiput(1150.00,580.17)(41.973,10.000){2}{\rule{0.970pt}{0.400pt}}
\put(309,147){\usebox{\plotpoint}}
\multiput(309.00,147.58)(1.009,0.499){137}{\rule{0.906pt}{0.120pt}}
\multiput(309.00,146.17)(139.120,70.000){2}{\rule{0.453pt}{0.400pt}}
\multiput(450.00,217.58)(0.657,0.499){163}{\rule{0.625pt}{0.120pt}}
\multiput(450.00,216.17)(107.702,83.000){2}{\rule{0.313pt}{0.400pt}}
\multiput(559.00,300.58)(0.677,0.498){65}{\rule{0.641pt}{0.120pt}}
\multiput(559.00,299.17)(44.669,34.000){2}{\rule{0.321pt}{0.400pt}}
\multiput(605.00,334.58)(0.894,0.499){155}{\rule{0.814pt}{0.120pt}}
\multiput(605.00,333.17)(139.311,79.000){2}{\rule{0.407pt}{0.400pt}}
\multiput(746.00,413.58)(1.116,0.498){95}{\rule{0.990pt}{0.120pt}}
\multiput(746.00,412.17)(106.946,49.000){2}{\rule{0.495pt}{0.400pt}}
\multiput(855.00,462.58)(1.918,0.492){21}{\rule{1.600pt}{0.119pt}}
\multiput(855.00,461.17)(41.679,12.000){2}{\rule{0.800pt}{0.400pt}}
\multiput(900.00,474.58)(1.178,0.499){117}{\rule{1.040pt}{0.120pt}}
\multiput(900.00,473.17)(138.841,60.000){2}{\rule{0.520pt}{0.400pt}}
\multiput(1041.00,534.58)(1.567,0.498){67}{\rule{1.346pt}{0.120pt}}
\multiput(1041.00,533.17)(106.207,35.000){2}{\rule{0.673pt}{0.400pt}}
\multiput(1150.00,569.58)(1.961,0.492){21}{\rule{1.633pt}{0.119pt}}
\multiput(1150.00,568.17)(42.610,12.000){2}{\rule{0.817pt}{0.400pt}}
\put(309,616){\usebox{\plotpoint}}
\multiput(309.00,616.58)(6.628,0.492){19}{\rule{5.227pt}{0.118pt}}
\multiput(309.00,615.17)(130.151,11.000){2}{\rule{2.614pt}{0.400pt}}
\multiput(450.00,627.59)(7.163,0.488){13}{\rule{5.550pt}{0.117pt}}
\multiput(450.00,626.17)(97.481,8.000){2}{\rule{2.775pt}{0.400pt}}
\multiput(559.00,635.59)(4.107,0.482){9}{\rule{3.167pt}{0.116pt}}
\multiput(559.00,634.17)(39.427,6.000){2}{\rule{1.583pt}{0.400pt}}
\multiput(605.00,641.58)(2.741,0.497){49}{\rule{2.269pt}{0.120pt}}
\multiput(605.00,640.17)(136.290,26.000){2}{\rule{1.135pt}{0.400pt}}
\multiput(746.00,667.58)(4.303,0.493){23}{\rule{3.454pt}{0.119pt}}
\multiput(746.00,666.17)(101.831,13.000){2}{\rule{1.727pt}{0.400pt}}
\multiput(900.00,678.92)(2.973,-0.496){45}{\rule{2.450pt}{0.120pt}}
\multiput(900.00,679.17)(135.915,-24.000){2}{\rule{1.225pt}{0.400pt}}
\multiput(1041.00,656.58)(2.117,0.497){49}{\rule{1.777pt}{0.120pt}}
\multiput(1041.00,655.17)(105.312,26.000){2}{\rule{0.888pt}{0.400pt}}
\multiput(1150.00,680.92)(0.795,-0.497){55}{\rule{0.734pt}{0.120pt}}
\multiput(1150.00,681.17)(44.476,-29.000){2}{\rule{0.367pt}{0.400pt}}
\put(855.0,680.0){\rule[-0.200pt]{10.840pt}{0.400pt}}
\end{picture}
\end{center}
\caption{Errors in $S$-wave phase shifts (in radians) vs the energy, computed with the theory using no contact terms (top line) and
the effective theory with leading-order (middle line) and next-to-leading order (bottom line). Figure taken from Ref.~\cite{lepage1997} with permission from Peter Lepage. }
\label{fig:lepage}
\end{figure}
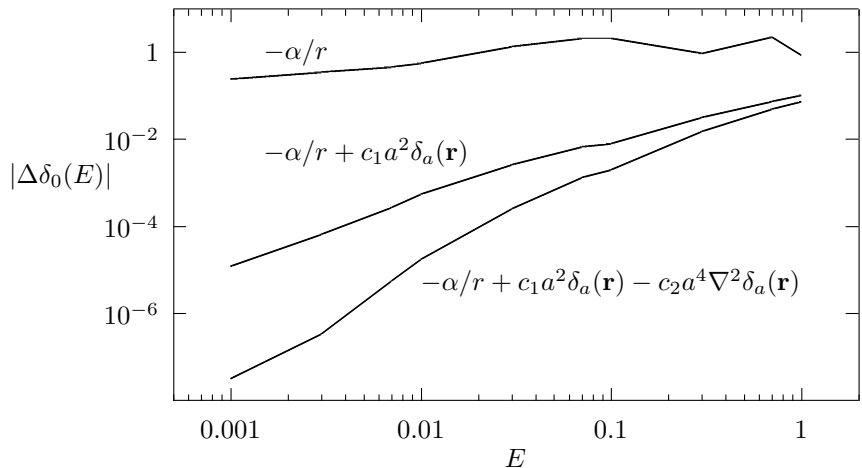

Lepage adjusted the low-energy constants $c_1$ and $c_2$ to data at low energies and then plotted the deviation $\Delta \delta_0(E)$ between $S$-wave phase shifts computed with the potential~(\ref{lepage}) and data. The results are reproduced in Fig.~\ref{fig:lepage}. Clearly, just using $V=V_{\rm long}=-\alpha/r$ is inaccurate at all energies. Adjusting $c_1$ to data (and setting $c_2=0$) yields a systematic improvement, particularly at low energies. The error is further decreased when $c_2$ is also adjusted. 

Figure~\ref{fig:lepage} is quite instructive. Clearly, the error grows with a power of the energy, i.e. $|\Delta\delta_0(E)|\propto E^n$. One can determine the   corresponding exponent $n$ at leading and next-to-leading order and compare with the expectations from the power counting of the potential~(\ref{lepage}). One can also infer the breakdown energy. 
(The reader is encouraged to do all this.) In summary, effective field theories provide one with a systematic tool to include unknown short-range physics into the description of low-energy phenomena.  

\subsection{Chiral effective field theory: highlights, status and challenges}  Chiral effective field theory employs pion exchange as the long-range part of the nuclear force and unknown short-range physics is captured by contacts. A highlight of this approach is that three-body forces enter at next-to-next-to leading order~\cite{vankolck1994,epelbaum2002}.  These consist of a long-range two-pion exchange (which was already introduced in the 1950s~\cite{fujita1957}), a one-pion-exchange contact (with low-energy constant $c_D$) and a three-nucleon contact (with low-energy constant $c_E$). 

Another highlight is that effective field theories provide one with a consistent approach to Hamiltonians and currents. To appreciate this one considers, for example, electromagnetic probes. Here, the charge current $\mathbf{j}$ couples to the electromagnetic field $\mathbf{A}$ in the usual way via $\mathbf{j}\cdot\mathbf{A}$. Of course, the current and the charge density $\rho$ also have to fulfill the continuity equation $\nabla\cdot\mathbf{j}+\partial_t\rho=0$. This requirement links currents to Hamiltonians via the Heisenberg equation of motion $\partial_t\rho=i[H,\rho]$. As the potential from effective field theory contains derivative operators, the commutator  $[H,\rho]$ can be a complicated expression. Nevertheless, effective field theory provides a consistent framework for relating currents to Hamiltonians. As a consequence of three-body forces, two-body currents are present, and the strengths of these currents also depend in a parameter-free way on the low-energy constants $c_D$ and $c_E$. The two-body currents provide us with noticeable corrections to magnetic moments and $\beta$-decay, as will be discussed in Section~\ref{subsec:quench}. 

It turned out, however, that the Weinberg power counting is not correct. Nogga, Timmermans, and van~Kolck showed that the theory is not properly renormalized at leading order~\cite{nogga2005}, because spurious bound states enter at sufficiently large cutoffs (in excess of 4~GeV) in attractive spin-triplet channels (in $P$ and $D$ waves), and the corresponding phase shifts are cutoff dependent. One can remedy this by promoting contacts that appear at higher order in the Weinberg power counting to leading order, but this approach seems ad hoc and the resulting power counting without a clear pattern. Most practitioners have stuck with the Weinberg power counting, probably also because overall accurate potentials only appear at next-to-next-to leading order, and cutoffs are usually kept closer to 500~MeV.  

It is interesting to look results obtained with the state-of-the-art nucleon-nucleon potential from chiral effective field theory by Reinert, Krebs and Epelbaum~\cite{reinert2018}. Those authors drove the expansion to a high order in the Weinberg power counting, and they  adjusted the low-energy constants to data for the nucleon-nucleon system. The resulting potential reproduced data with increasing accuracy and precision as the order was increased. Three-body forces were adjusted by the LENPIC collaboration in Ref.~\cite{epelbaum2019}. Here, the accuracy and precision was not as high as in the nucleon-nucleon sector. Finally, the LENPIC collaboration used these nucleon-nucleon and three-nucleon potentials and computed light nuclei~\cite{maris2021}. Those results revealed the following. Computations that only employed the leading-order potential yielded nuclei that are unstable with respect to emission of $\alpha$-particles. The ground-state energy of the nucleus $^6$Li, for instance, was above the $\alpha+d$ threshold. Similar problems occurred $^6$He, $^{10}$Be, $^{10}$B, and $^{12}$C. However, nuclear binding of light nuclei was  accurately described at higher orders. The follow-up study~\cite{maris2022} then revealed that -- even at high order -- $^{40,48}$Ca were overbound by about 1~MeV per nucleon and that the corresponding radii were about 20\% too small.  

These problems -- overbinding and too small radii -- were known for some time. To overcome that problem, Ekstr\"om et al.~\cite{ekstrom2015a} adjusted a chiral interaction at next-to-next-to-leading order to data in two- and three-nucleon systems and also used the binding energies and charge radii (where measured) of $^4$He, $^{14}$C and $^{16,22,24,25}$O in the calibration. The resulting potential NNLO$_{\rm sat}$ exhibits accurate charge radii and binding energies for light to medium-mass nuclei.

An alternative approach to the saturation problem was taken in lattice nuclear effective field theory~\cite{lee2009}. There, the practitioners introduced two additional three-body contacts and adjusted the strengths of three-body forces to $\alpha-\alpha$ scattering (an eight-body process)~\cite{elhatisari2015} and to the triple-$\alpha$ resonance (Hoyle state) in $^{12}$C and to binding energies of 14 nuclei from $^4$He to $^{40}$Ca. As a result, charge radii were accurate for light to medium-mass nuclei~\cite{elhatisari2024}.

The discussion of the last three paragraphs shows that nuclear binding and saturation (i.e. accurate binding energies and nuclear densities) are finely tuned. Even precision forces have failed to accurately saturate nuclei if their calibration was limited to only $A=2,3$ nuclei. Practical solutions consist of employing critical data in heavier nuclei in the calibration. 

\subsection{Renormalization group transformations generate softer interactions} 
Equation~(\ref{spsize}) below shows that the number of single-particle states required to compute a nucleus increases with the third power of the momentum cutoff. This makes using low-momentum interactions an attractive proposition for many-body computations. One way is, of course, to simply use a lower cutoff when adjusting the low-energy constants of a chiral potential to data, i.e. to directly construct an interaction with a lower cutoff. Interestingly there are also tools from the renormalization group that allow one to decrease the cutoff for any potential whose cutoff might be too large for a convenient many-body computation. I briefly review two such approaches; for reviews see Refs.~\cite{bogner2010,furnstahl2013}. 

In 2003 Bogner, Kuo and Schwenk proposed to use renormalization group transformations to integrate out high-momentum modes from a nucleon-nucleon interaction that has a high cutoff~\cite{bogner2003}. One requires that the on-shell $T$-matrix remains invariant (for momenta below the cutoff) as the cutoff is lowered. This yields an evolution equation whose solution is the low-momentum nucleon-nucleon potential $V_{\rm{low}-k}$. At sufficiently low cutoffs (around $\lambda=1.8$ to $2.0$~fm$^{-1}$), the potential $V_{\rm{low}-k}$ becomes almost independent from the nucleon-nucleon interaction one started from; its matrix elements vanish for momenta above the cutoff. However, a renormalization-group transformation also induces three-nucleon forces and the evolution equation of the three-body $T$ matrix is more complicated. Instead, the practice was to adjust the three-body contacts $c_D$ and $c_E$ at a give cutoff to data in $A=3,4$ systems~\cite{nogga2004}, and thereby to combine the leading three-nucleon potential from chiral effective field theory with $V_{\rm{low}-k}$. 

In 2007, Bogner, Furnstahl and Perry proposed an alternative method that has been popular ever since~\cite{bogner2007}. They applied ideas from White~\cite{white2002}, Glazcek and Wilson~\cite{glazek1993}, and Wegner~\cite{wegner1994} to nuclear interactions. They used a similarity transformation
\begin{equation}
\label{srg}
H(s)= U(s)H U^\dagger(s) \ , 
\end{equation}
where $U(s)$ is a unitary operator to decouple low and high-momentum modes in $H(s)$. 
Taking the derivative of Eq.~(\ref{srg}) with respect to $s$ then yields the evolution equation
\begin{equation}
    \frac{d H(s)}{ds} = \left[\eta(s),H(s)\right] \ , 
\end{equation}
where 
\begin{equation}
\eta(s)=\frac{dU(s)}{ds}U^\dagger(s)
\end{equation}
and one has to solve a set of ordinary differential equations for the matrix elements of $H(s)$ giving a starting Hamiltonian as the initial value. One can, of course,  apply the same similarity transformation to three-body forces. 

The question now arises which generator to use in the similarity renormalization group transformation. One popular choice has been  
\begin{equation}
\label{SRGgen}
    \eta(s) = \left[T_{\rm kin},H(s)\right] \ ,
\end{equation}
where $T_{\rm kin}$ is the kinetic energy. It is then clear that the evolution stops when the potential $V_s(k',k)$ becomes diagonal in momentum space. In practice, one only needs to achieve a band-diagonal structure to decouple low and high-momentum modes. This is because band-diagonal matrices have localized eigenvectors. In any case, the generator~(\ref{SRGgen}) drives a given potential $V_{s=0}(k',k)$ in momentum space close to a diagonal matrix form.  

\begin{figure}[htb]
\centering
\includegraphics[width=0.49\textwidth]{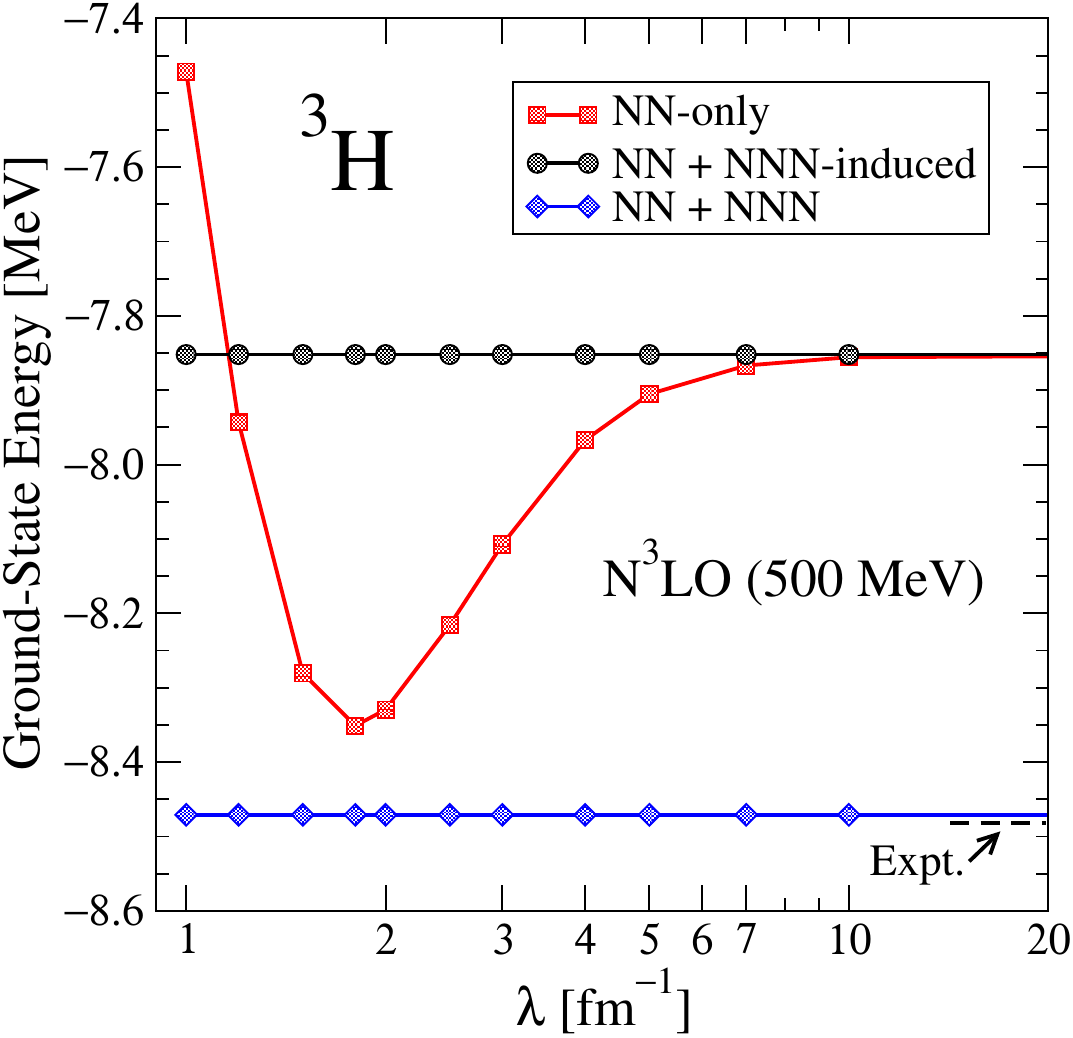}
\includegraphics[width=0.49\textwidth]{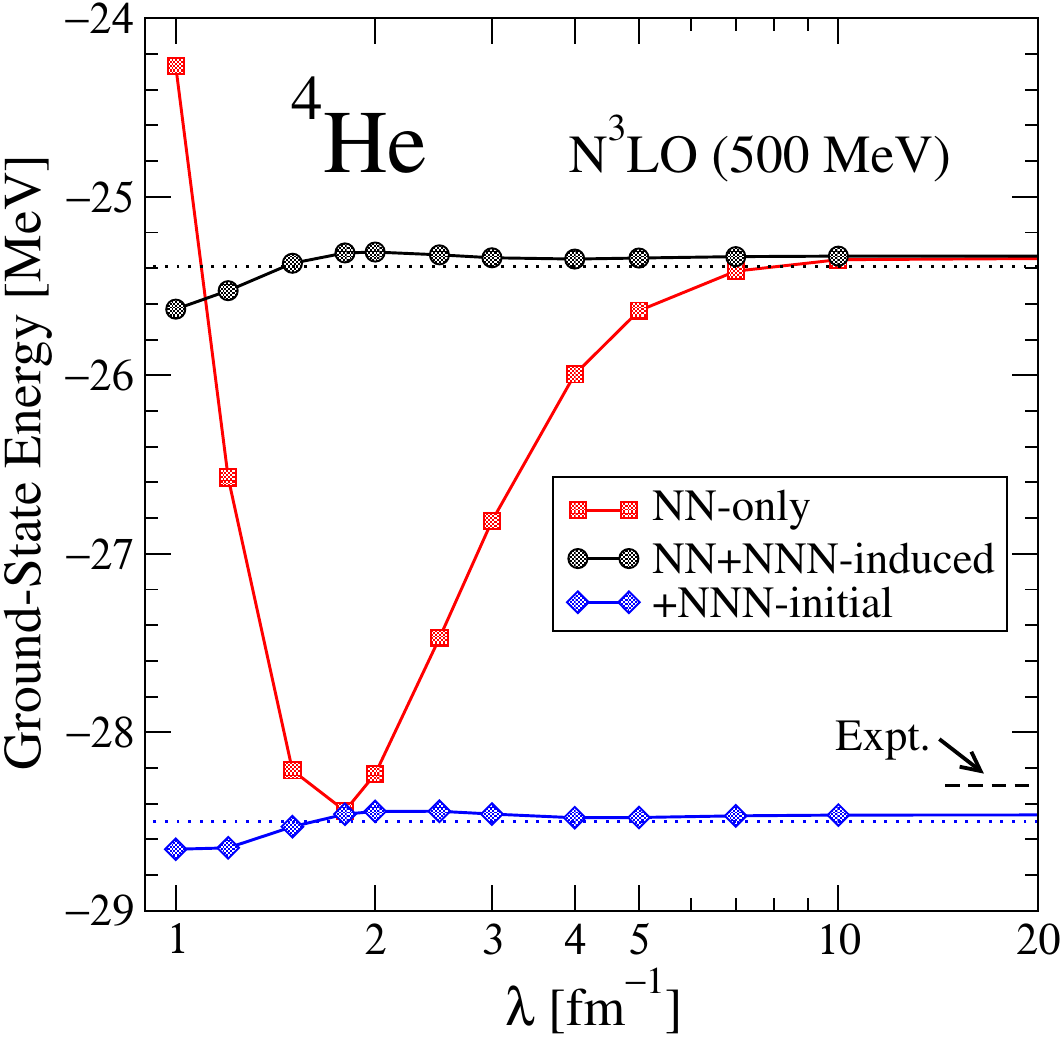}
\caption{Ground-state energies for $^3$H (left) and $^4$He (right) as a function of the effective cutoff $\lambda$ in the similarity renormalization group. Results are shown for an evolution of a nucleon-nucleon potential only (NN-only), for an evolution that starts with nucleon-nucleon forces and also generates and evolves the induced three-nucleon potential (NN+NNN induced) and for an evolution that starts from two-and three-nucleon forces and evolves them simultaneously (NN+NNN). Taken from arXiv:0905.1873 with permission from the authors, see also Ref.~\cite{jurgenson2009}.}
\label{fig:jurgenson}
\end{figure}

It is instructive to compare the similarity renormalization group transformation and the $V_{\rm{low}-k}$ potentials. The former results from a unitary transformation and therefore is phase-shift equivalent to the starting potential. In contrast, the latter is phase-shift equivalent only up to the cutoff. Another advantage is that the consistent construction of three-nucleon forces via the similarity renormalization group is technically simpler than for $V_{\rm{low}-k}$. 
Jurgenson, Furnstahl and Navr{\'a}til~\cite{jurgenson2009} 
presented the simultaneous evolution of nucleon-nucleon and three-nucleon forces and studied the impact in light nuclei. Figure~\ref{fig:jurgenson} shows their results for $^3$H and $^4$He. Here, the evolution is shown with respect to the effective momentum cutoff $\lambda\equiv s^{-1/4}$. 

The figure for each nucleus is best read from right to left; the results for the unevolved initial potential are obtained for $\lambda\to\infty$.  For the triton, the ground-state energy is cutoff independent if one evolves the nucleon-nucleon and three-nucleon potential. This is true independent of whether the initial potential contains a three-nucleon force (NN+NNN) or not (NN+NNN induced). In contrast, if one only involves a nucleon-nucleon potential and neglects any three-nucleon potentials that are generated in the evolution (NN only), the ground-state energy becomes cutoff dependent. The results are similar in $^4$He, though results become cutoff dependent at sufficiently small cutoffs even for the cases (NN+NNN induced) and (NN+NNN). (The reader is encouraged to think about what is missing.)   

Let me finally introduce another popular interaction. Hebeler et al.~\cite{hebeler2011} started from a high-precision nucleon-nucleon potential by Entem and Machleidt~\cite{entem2003} and used the similarity renormalization group transformation to lower the cutoff. Instead of evolving three-nucleon forces they took the leading three-nucleon forces from chiral effective field theory and adjusted the low-energy constants $c_D$ and $c_E$ to the triton binding energy and the charge radius of $^4$He. One of the resulting interactions, named 1.8/2.0(EM) because it was evolved to $\lambda=1.8$~fm$^{-1}$ and used a cutoff of 2.0~fm$^{-1}$ for the three-body potential, has turned out to be very accurate for energies of ground and excited states.

\section{Single-particle bases}
\label{sec:sp}
\subsection{How many single-particle states does it take to model a nucleus?}
\label{sec:howmany}
The coupled-cluster, Gorkov-Green's functions, IMSRG, and nuclear lattice effective field theory all start from a discrete single-particle basis with states $|q\rangle$. As an example one considers nuclear lattice effective field theory~\cite{laehde2019}, where $q=(\mathbf{r}_q, s_q, \tau_q)$ labels the discrete position, spin and isospin projections. The method employs a three-dimensional lattice with spacing $a_0$ and extent $L$ (where $L/a_0$ is an integer of order ten) in position space. Clearly, to describe a nucleus with a radius $R$ one must have $L>2R$. Interactions from quantum field theory employ a momentum cutoff $\Lambda$ and one has $\Lambda\sim a_0^{-1}$. One can use phase-space arguments to find the number of single-particle states ${\cal M}$ required in the computation, i.e. required to compute a nucleus with radius $R$ from an interaction with momentum cutoff $\Lambda$. This yields 
\begin{equation}
\label{spsize}
    {\cal M}\equiv g (R/a_0)^3\sim g \Lambda^3 R^3.  
\end{equation}    
Here, $g$ is the spin-isopsin degeneracy. One sees that ${\cal M}$ is cubic in the cutoff. Thus, much is to be gained from lowering the momentum cutoff, and this explains the popularity of methods~\cite{bogner2003,bogner2007} that accomplish this. As nuclei are saturated systems, $R\sim A^{1/3}$, with a prefactor of the order of 1~fm. One sees that weakly bound nuclei require more effort because extended nuclear halos increase the radius $R$. The simple expression~(\ref{spsize}) also makes clear that heavier nuclei require larger model spaces for the same interaction, i.e. the same momentum cutoff $\Lambda$.

The harmonic oscillator is a very popular basis as well. Here, single-particle states $|q\rangle$ are labeled by quantum  numbers $q=(n_q,l_q,j_q,{j_z}_q,\tau_q)$ using the principle quantum number, orbital angular momentum, total angular momentum, its projection, and the isospin projection. Let $m$ denote the nucleon mass. For a single-particle basis with oscillator spacing $\hbar\omega$ and oscillator length $b=\sqrt{\hbar/(m\omega)}$ consisting of spherical shells with up to $(N_{\rm max}+3/2)\hbar\omega$ of excitation energy, $L\sim N_{\rm max}^{1/2}b$ and $\Lambda\sim N_{\rm max}^{1/2}b^{-1}$;  more accurate expressions are presented in Refs.~\cite{furnstahl2014,more2013,konig2014}. Thus, the formula~(\ref{spsize}) derived for the lattice also applies  to the spherical harmonic oscillator basis once these identifications have been made. Present ab initio computations of nuclei (where $g=4$) employ lattices (or harmonic oscillator bases) with ${\cal M}\approx 4\times 10^3$~\cite{lu2019,hagen2022}.

\subsection{Hamiltonian matrix elements and the center of mass}
\label{sec:matele}

The Hamiltonians of interest are of the form 
\begin{equation}
\label{ham}
    \hat{H}= T_{\rm kin}-T_{\rm CoM} +V_{NN} +V_{NNN} \ .
\end{equation}
Here, $T_{\rm kin}$ denotes the kinetic energy of the $A$-body system, $T_{\rm CoM}$ is the kinetic energy of the center of mass, and $V_{NN}$ and $V_{NNN}$ denote nucleon-nucleon and three-nucleon potentials. 
Potentials from chiral effective field theory are often written down in momentum space~\cite{epelbaum2009,machleidt2011,hammer2020} and one has $V_{NN}=V_{NN}(p';p)$ and $V_{NNN}=V_{NNN}(p',q';p,q)$ in terms of relative (Jacobi) momenta $p$ and $q$;  the notation of spin and isospin degrees of freedom are suppressed. On position-space lattices such potentials can be computed efficiently via fast Fourier transformations~\cite{elhatisari2016}, or via analytical means~\cite{piarulli2015,piarulli2016}. Importantly, the potential is short-ranged in position space which leads to sparse matrices on the corresponding lattice.  
The computation of these matrix elements in the harmonic oscillator is solved via numerical integration. The  transformation from the center-of-mass  to the laboratory system is a tedious (yet solved) step. It involves the Brody-Moshinsky transformation for two-nucleon potentials  and a double Brody-Moshinsky transformation for three-nucleon potentials~\cite{navratil2007}. The resulting matrices are block diagonal (because of quantum numbers such as particle number, parity, and angular momentum) but not anymore sparse, because the short-range structure of the interaction is not anymore expressed in the harmonic-oscillator basis and the Galilean invariance is broken. This results in large demands for storage of such matrix elements, particularly for the three-nucleon potentials. Hebeler et al.~\cite{hebeler2015b} and Miyagi et al.~\cite{takayuki2022} have devised clever schemes to deal with this problem, see also Ref.~\cite{hebeler2023}. The main insight is that one only needs a small fraction of the three-nucleon matrix elements for calculations in the normal-ordered two-body approximation (see Sect.~\ref{sec:no} for details). For a recent review see Ref.~\cite{hebeler2021}. 

The Hamiltonian~(\ref{ham}) employs the intrinsic kinetic energy $T_{\rm kin}-T_{\rm CoM}$ and therefore neither contains the center-of-mass coordinate nor its conjugate momentum. Thus, it only depends on $3(A-1)$ degrees of freedom. The model space, however, is based on single-particle states in the laboratory system and does not remove the center-of-mass degree of freedom. The Hartree-Fock state, for instance, is a product of $A$ single-particle states. Hagen, Papenbrock and Dean~\cite{hagen2009a} found that coupled-cluster wave functions  (which are also based in the laboratory system) factorize to a very good approximation into an intrinsic wave function and a Gaussian center-of-mass state. Jansen~\cite{jansen2012} found similar results for coupled-cluster computations of excited states, and the factorization was also confirmed in IMSRG computations~\cite{parzuchowski2017}. Calculations show that the approximate factorization even holds for Hartree-Fock states. Thus, using an intrinsic Hamiltonian in a sufficiently large model space yields approximate eigenstates that (to a very good approximation) respect the  translational invariance of the Hamiltonian; it is not necessary to require that the basis exhibits that symmetry as well. Clearly, the Hamiltonian knows best, and all one has to provide it with is a sufficiently large (i.e. low-energy complete) model space.      

The challenges discussed in this Section, i.e. computing matrix elements in the laboratory system and dealing with the center of mass, can of course be avoided when using Jacobi coordinates that remove the center of mass. However, the proper antisymmetrization of the nuclear wave function becomes so cumbersome in those coordinates that their application is limited to light nuclei up to mass $A\lesssim 7$ or so~\cite{barnea1998,bacca2004}. The center-of-mass problem is completely avoided in the no-core shell model, because that employs an $A$-body model space where the center of mass is always in the ground state of the harmonic oscillator basis~\cite{barrett2013}.

\section{The mean field points the way}
\label{sec:hf}
For wavefunction based methods, the mean-field calculation is the most important step in computing nuclear properties. While the mean-field energy is rather inaccurate, the mean field is key to organizing and including correlations into the nuclear wave function, and to determine a nucleus's shape and size. Furthermore, mean-field states that break symmetries associated with conservation of angular momentum or particle numbers point to physical concepts such as nuclear deformation and superfluidity. In these cases the mean field state already indicates that the corresponding low-energy excitations are rotational and pairing rotational bands, respectively. This connection between mean field and low-lying excitations is nicely highlighted in the textbook~\cite{altland2006}.  For most of this lecture I will assume that the mean field results from a Hartree-Fock (rather than a Hartree-Fock-Bogoliubov) calculation. The main reason is that coupled-cluster theory includes pairing correlations on top of the mean-field state.   

Hartree-Fock calculations yield the reference state
\begin{equation}
\label{refket}
    |\psi_0\rangle \equiv \prod_{i=1}^A \hat{a}_i^\dagger |0\rangle \ .
\end{equation}
Here $|0\rangle$ denotes the vacuum and $a_i^\dagger$ creates a nucleon in the single-particle state $|i\rangle\equiv a_i^\dagger|0\rangle$. The label $i=(n_i,\pi_i,j_i, {j_z}_i, {\tau_z}_i)$ denotes the principal shell $n_i$, parity $\pi_i$, angular momentum $j_i$, its $z$-projection ${j_z}_i$, and the isospin projection ${\tau_z}_i$ in a spherical Hartree-Fock basis. In a deformed (axially symmetric) basis one would use $i=(\alpha_i,\pi_i,{j_z}_i, {\tau_z}_i)$  and let $\alpha_i$ be a quantum number needed for further distinction of Nilsson-like orbitals. 
It is customary to use the subscripts $i,j,k,\ldots$ to denote single-particle states occupied in the reference state $|\psi_0\rangle$ and label unoccupied states as $a,b,c,\ldots$. Thus $|i\rangle$ and $|a\rangle$ are hole and particle states, respectively. One uses subscripts $p,q,r,\ldots$ when no distinction between hole and particle states is made.

\subsection{Fermi momentum as a dividing scale}
\label{sec:dividing}
The Hartree-Fock calculation introduces the Fermi momentum $k_F$ as a relevant scale which sets the inter-nucleon distance ($\propto 1/k_F$) and the density ($\propto k_F^3$) of the nucleus. This allows one to view $k_F$ as a dividing scale. Momenta smaller and larger than $k_F$ correspond to long-range and short-range physics, respectively.

\begin{figure}[htb]
\centering
\includegraphics[width=0.6\textwidth]{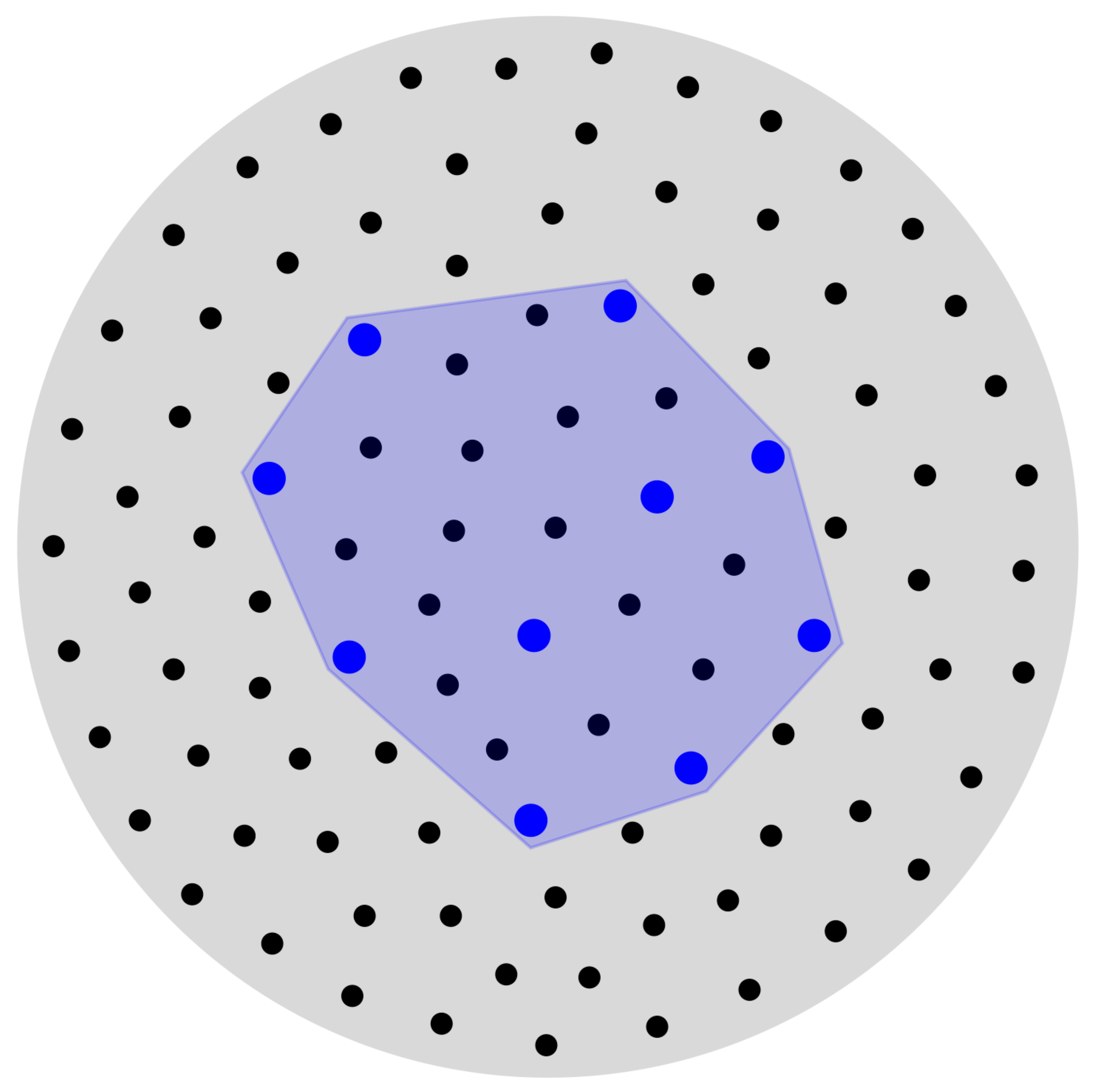}
\caption{Position-space cartoon of a localized single-particle basis in two dimensions resulting from re-localizing Hartree-Fock states. The blue points represent the centers of localized hole states while the black points represent the centers of localized particle states. The area of a two-dimensional nucleus (light blue) and its complement (gray) are also shown. }
\label{fig:localized}
\end{figure}

To see how $k_F$ emerges in a finite nucleus (without invoking the free Fermi gas) one considers the following arguments.
Hartree-Fock calculations partition the single-particle basis into hole states $|i\rangle$ and particle states $|a\rangle$ such that the Hartree-Fock Hamiltonian 
\begin{equation}
    H_{\rm HF} \equiv \sum_{pq} f^q_p \hat{a}^\dagger_q \hat{a}_p
\end{equation}
is diagonal, i.e. the matrix elements fulfill $f_p^q=\delta_p^q f_p^p$, where
\begin{equation}
\label{fockmat}
    f_p^q \equiv \langle q|H|p\rangle +{1\over 2}\sum_i \langle qi|H|pi\rangle + {1\over 2} \sum_{ij} \langle qij|H|pij\rangle  \ .
\end{equation}
On the one hand, this basis is attractive because it allows us to interpret $f_p^p$ as single-particle energies. On the other hand, this basis is unattractive because its states $|p\rangle$ are localized in energy but delocalized in position space.  However, one can make a Gedankenexperiment and perform a unitary transformation $U(A)$ among the hole states and an independent unitary transformation $U({\cal M}-A)$ of the particle states without changing the energy~(\ref{Eref}) of the reference state. Such transformations then 
produce localized basis states 
\begin{equation}
\label{localized}
    |q\rangle = |\mathbf{r}_q, s_q,\tau_q\rangle \ , 
\end{equation}
where $(\mathbf{r}_q, s_q,\tau_q)$ denote the position, spin and isospin projections\footnote{This has been accomplished in quantum chemistry~\cite{hovik2012} but not yet tried out in nuclear physics.}. 
In what follows I will assume that one has arrived at such a localized basis. This will allow me to employ arguments based on the short-range nature of the nuclear interaction. A cartoon (for a two-dimensional nucleus) is depicted in Fig.~\ref{fig:localized}. 
The key to appreciate is that there are particle states also ``inside'' the volume of the nucleus.

The nearest-neighbor distance between hole states is proportional to the inverse Fermi momentum, $k_F^{-1}$, while the nearest-neighbor distance between particle states is proportional to the inverse  momentum cutoff, $\Lambda^{-1}$. For estimates one notes that the inter-nucleon distance at nuclear saturation density $\rho_0\approx 0.16$~fm$^{-3}$ is $\rho_0^{-1/3}\approx 1.8$~fm. However, if one takes spin-isospin symmetry into account (so that four nucleons can be at the same localized point), the distance between the blue points in Fig.~\ref{fig:localized} becomes $(\rho_0/4)^{-1/3}\approx 2.9$~fm, and this is at or somewhat above the range of nuclear interactions. 

It is then clear that particle-hole excitations, generated by short-range interactions, will mainly introduce short-range correlations, with the result of getting the structure of $\alpha$-particle clusters (and the interactions between them) right. This picture is consistent with the insights gained from nuclear lattice effective theory about the importance of $\alpha$-clustering and the interactions between such clusters~\cite{elhatisari2016}.

\subsection{The energy of the reference state is proportional to the mass number}
\label{sec:EHF}

The energy of the reference state is 
\begin{eqnarray}
\label{Eref}
    E_{\rm ref} &\equiv& \langle \psi_0|H|\psi_0\rangle \nonumber\\
    &=&\sum_i\langle i|H|i\rangle +{1\over 2}\sum_{ij} \langle ij|H|ij\rangle + {1\over 6}\sum_{ijk} \langle ijk|H|ijk\rangle\ \ .
\end{eqnarray}
The sums run over all hole states. One now assumes that one deals with the localized basis states~(\ref{localized}).  (Localized hole states are depicted by dots that are at their center in Fig.~\ref{fig:localized}). As the nuclear interaction is short ranged, for fixed $i$ the states $j$ (and $k$) must be close to $i$ (and to each other), i.e. $|\mathbf{r}_i-\mathbf{r}_j|$ must be smaller than the range of nuclear interaction. Thus, the energy of the reference state is proportional to $A$. This is an important result. It implies that the Hartree-Fock method is size extensive. 

This makes it interesting to return to the Hartree-Fock basis where the Fock matrix~(\ref{fockmat}) is diagonal. The Hartree-Fock energy is presented in Eq.~(\ref{Eref}). 
Assuming that there are no tricky cancellations, each of the three sums in this equation must scale proportional to the mass  number $A$. This allows one to understand the scaling of the single-particle energies $f_i^i$ in Eq.~(\ref{fockmat}). Using the empirical shell model one finds
\begin{eqnarray}
\label{shellgap}
    \sum_i f_i^i &=& \sum_s n_s \varepsilon_s \nonumber\\
    &\propto& A^{2/3} \sum_s \varepsilon_s\nonumber\\
    &\propto& -A^{2/3} \Delta \varepsilon \sum_s s\nonumber\\
    &\propto& - A^{4/3} \Delta \varepsilon \ .
\end{eqnarray}
Here, the sum over $s$ is over major shells (and their total number is proportional to $A^{1/3}$), and  $n_s\propto A^{2/3}$ is the occupation number of the shell $s$, and the corresponding single particle energies are equally spaced and denoted as $\varepsilon_s=-s\Delta \varepsilon$, where $\Delta\varepsilon$ is the energy gap between shells. One thus has 
\begin{equation}
\label{scaleSPE}
    \Delta \varepsilon \approx \frac{k_F^2}{2m} A^{-1/3}  \ .
\end{equation} 
Here, the proportionality constant is approximately the Fermi energy $E_F=k_F^2/(2m)\approx 40$~MeV with $m$ denoting the nucleon mass. 

The average sizes of the diagonal two-body and three-body matrix elements then are
\begin{equation}
\begin{aligned}
\label{scaleME}
    \langle ij|H|ij\rangle &\propto A^{-1} \ , \\
    \langle ijk|H|ijk\rangle &\propto A^{-2} \ .
\end{aligned}
\end{equation} 
These relations can also be derived under the assumption that the nuclear interaction is short ranged and that the Hartree-Fock wave functions $\langle\mathbf{r}|i\rangle$ are delocalized over a spherical region with radius proportional to $A^{1/3}$ (and thus having amplitudes $\langle\mathbf{r}|i\rangle\propto A^{-1/2}$).

These arguments suggest that one-, two, and three-body matrix elements exhibit a separation of scales in heavy nuclei. This can be exploited in the computation of excited states, see Sect.~\ref{sec:1p1h}.

\subsection{The advantages of breaking symmetries}
The nuclear shell model identifies proton and neutron numbers 2, 8, 20, 28, 50, 82, and 126 as closed shells~\cite{mayer1955}. Most nuclei thus have open shell(s). For such nuclei one can not write down a product state with zero angular momentum. Using a single-particle basis with good angular momentum ($j$) and angular-momentum projection ($j_z$) the best one can do  (for even-even nuclei) is to write down a reference state where pairs of time-reversed orbitals (i.e. pairs of orbitals with $\pm j_z$ but all other quantum numbers equal) are occupied. Such a product state is invariant under time reversal, has zero total angular momentum projection, but lacks good total angular momentum. It is a deformed yet axially symmetric state which breaks rotational symmetry. This has two important consequences. 

\begin{figure}[htb]
\centering
\includegraphics[width=0.9\textwidth]{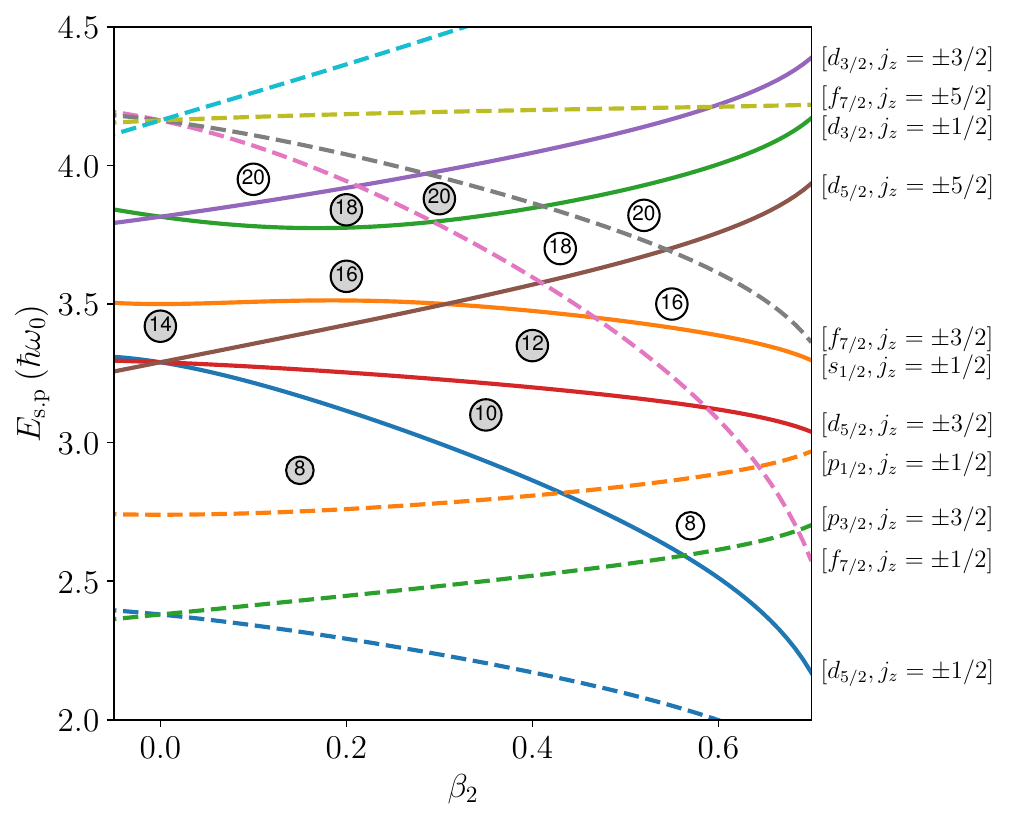}
\caption{The Nilsson diagram for light nuclei shows energies of single-particle states as a function of the  quadrupole deformation $\beta_2$. States with negative (positive) parity are shown as dashed (full) lines, and numbers in circles indicate shell closures. Taken from Ref.~\cite{sun2024b} with permission from the authors.}
\label{fig:nilsson}
\end{figure}

First, the Nilsson model now becomes the appropriate shell model~\cite{nilsson1955}, and for many nuclei one can find a deformation such that there is a significant shell gap at the Fermi surface. Then one again deals with closed-shell nuclei -- albeit in a more generalized sense. In practice, such a deformed reference state can be obtained from Hartree-Fock computation that starts from a state where (spherical) harmonic-oscillator states are occupied according to the Nilsson model. This allows one to apply single-reference methods, e.g. many-body perturbation theory or coupled-cluster theory. An example is shown in Fig.~\ref{fig:nilsson}. 

Second, one needs to discuss symmetry restoration. 
A deformed reference state is not unique because any rotation of it yields another state with the same Hartree-Fock energy. Thus one deals with a ``multi-reference'' situation where different reference states are connected by a symmetry operation. It is well known how to restore symmetries of such states~\cite{sheikh2021}. The important point is that symmetry restoration yields an energy gain that is not size extensive, and in fact decreases with increasing mass number. Thus, it contributes very little to the bulk of the binding energy. This is because symmetry restoration or projection yields rotational bands, and the corresponding energy scale decreases with increasing mass number and is about the lowest energy observed in nuclei. Thus, this is really infrared physics, and it is is well separated in energy from other phenomena. (The rotational energy scale is less than 100~keV in rare earth nuclei and tens of keV in actinides.) This also allows one to describe and understand them in terms of collective models~\cite{rowe2010,bohr1975} and effective theories~\cite{papenbrock2011,coelloperez2015,chen2017,papenbrock2020}.   

While the previous discussion focused on the breaking of spherical symmetry, Hartree-Fock-Bogoliubov computations also break proton and neutron numbers to properly capture superfluidity via the mean-field. If one stops at the mean-field description of nuclei, this symmetry breaking (and the ensuing symmetry restoration) is essential in medium-mass and heavy nuclei. Pairing rotations result as the generalized excited states upon the restoration of particle numbers~\cite{brink-broglia2005}, and -- like rotational bands in deformed nuclei -- these are low-lying excitations and describe infrared properties~\cite{papenbrock2022}. In contrast to rotations (which include $A$-particle--$A$-hole excitations), however, pairing correlations are dominated by two-particle--two-hole excitations. Thus, they can also be included by post Hartree-Fock methods such as coupled-cluster theory. 

Summarizing, there are two big advantages of breaking symmetries in the mean field. The first is that the symmetry breaking reveals important physics such as nuclear deformation (in the case of breaking rotational symmetry) and superfluidity (in the case of breaking particle numbers). The second is that the restoration of symmetries is about including  long-wavelength physics into the nuclear ground state. It is universal because it leads to rotational and pairing rotational bands in the case of breaking rotational symmetry and particle numbers, respectively. These infrared phenomena are separated in scale from the short-range correlations that make the bulk of nuclear binding. Thus, one can use different methods to include them. This separtion of scales has been exploited recently in ab initio computations~\cite{sun2024}.   

\subsection{What are a nucleus's references?}
Nuclei are complex systems, and a nucleus can exhibit various structures. The nucleus $^{48}$Ca, for instance exhibits a spherical ground state, a $0^+$ state at 3.35~MeV that is the head of a  deformed band, and another $0^+$ state at 5.21~MeV that is the head of a superdeformed rotational band~\cite{ideguchi2022}. The construction of deformed reference states can be based on Nilsson diagrams~\cite{nilsson1955}, see Fig.~\ref{fig:nilsson} as an example. Alternatively, one can compute the corresponding reference states by adding a quadrupole constraint to the Hamiltonian and compute the Hartree-Fock state at a given deformation.  Good reference states are in well-localized energy minima; they can serve as starting points for single-reference calculations with subsequent symmetry projections. Other examples are odd-mass deformed nuclei, where different spin and parity single-particle states of the odd nucleon in the (symmetry-broken) reference result in different rotational bands after symmetry projection. One sees again that heuristics and the mean-field calculation are essential in providing us with a starting point for spectra in nuclei. One can then include few-particle--few-hole correlations to capture the bulk of the binding energy and the corresponding short-range correlations, and symmetry projection to project out rotational bands associated with the reference state. This approach is both physically and technically appealing because it simplifies the understanding and computation. A beautiful example for how well this physical picture works was given by \cite{ropke1998} who were able to group most of the low-lying states in the odd-odd deformed nucleus $^{26}$Al into rotational bands. Ab initio computations are following this path as well~\cite{sun2024b}.    

\begin{figure}[htb]
\centering
\includegraphics[width=0.80\textwidth]{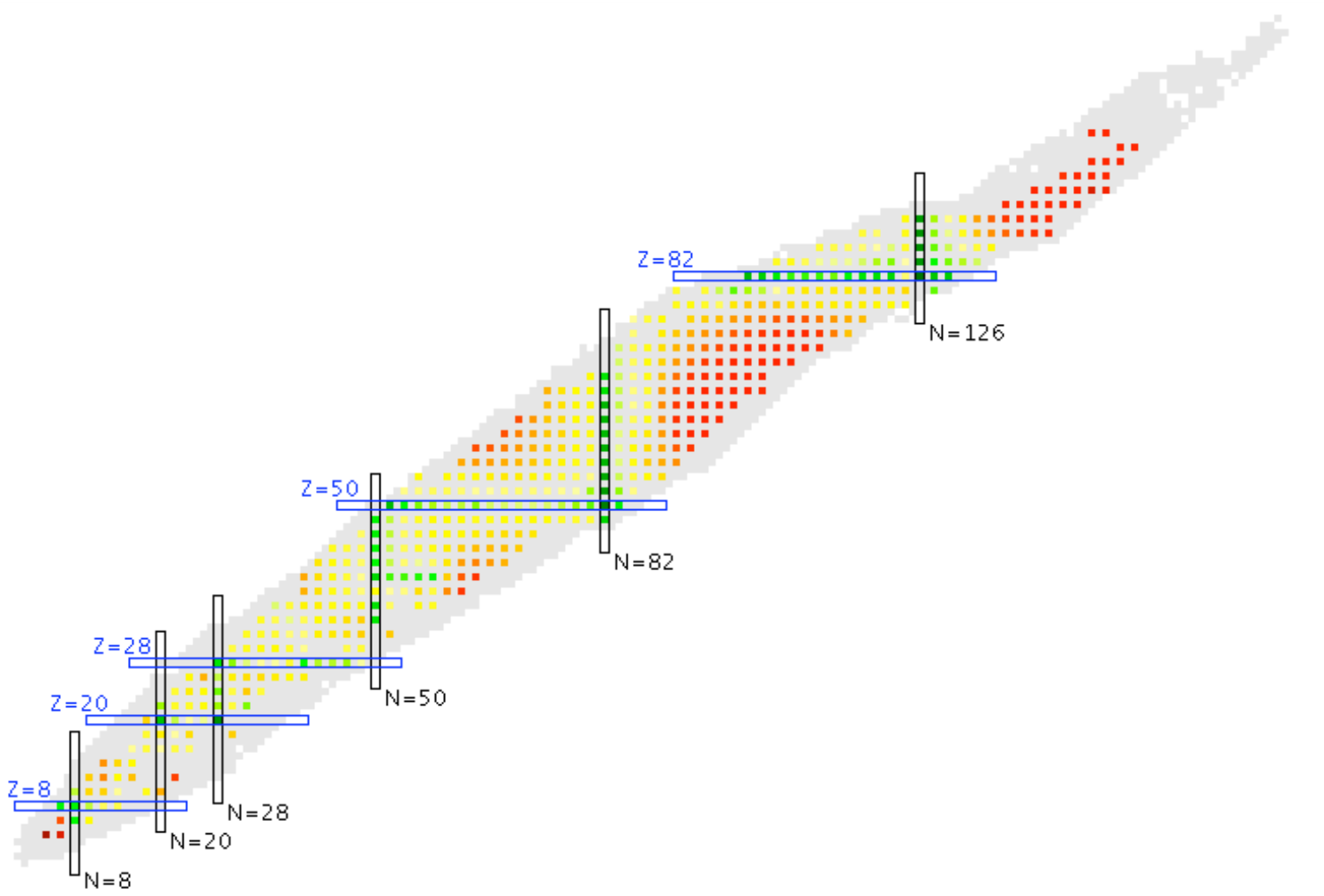}
\caption{The ratio $E(4^+)/E(2^+)$ for even-even nuclei across the nuclear chart. The red color shows nuclei for which this ratio is close to that of a rigid rotor, i.e. $E(4^+)/E(2^+)\approx 3.3$. These nuclei exhibit rigid deformation. Also shown are magic numbers with neutron numbers $N$ and charge $Z$ as indicated. Screenshot taken from the National Nuclear Data Center at https://www.nndc.bnl.gov/nudat3/.}
\label{fig:chart}
\end{figure}

The question then arises about how which nuclei have ``good'' reference states, i.e. either a single reference state or a family of equivalent reference states that are connected by symmetry operations such as rotations? Inspection of the nuclear chart in Fig.~\ref{fig:chart} shows that there are many candidates. First are all doubly magic nuclei, shown at the intersections of magic numbers; they exhibit spherical Hartree-Fock references. Second are all even-mass semi-magic nuclei; these have spherical Hartree-Fock-Bogoliubov reference states. Finally there is a large number of nuclei with rigid deformation (shown in red) which have good axially-symmetric reference states.   

\subsection{Natural orbitals and localization}
\begin{figure}[htb]
\centering
\includegraphics[width=0.50\textwidth]{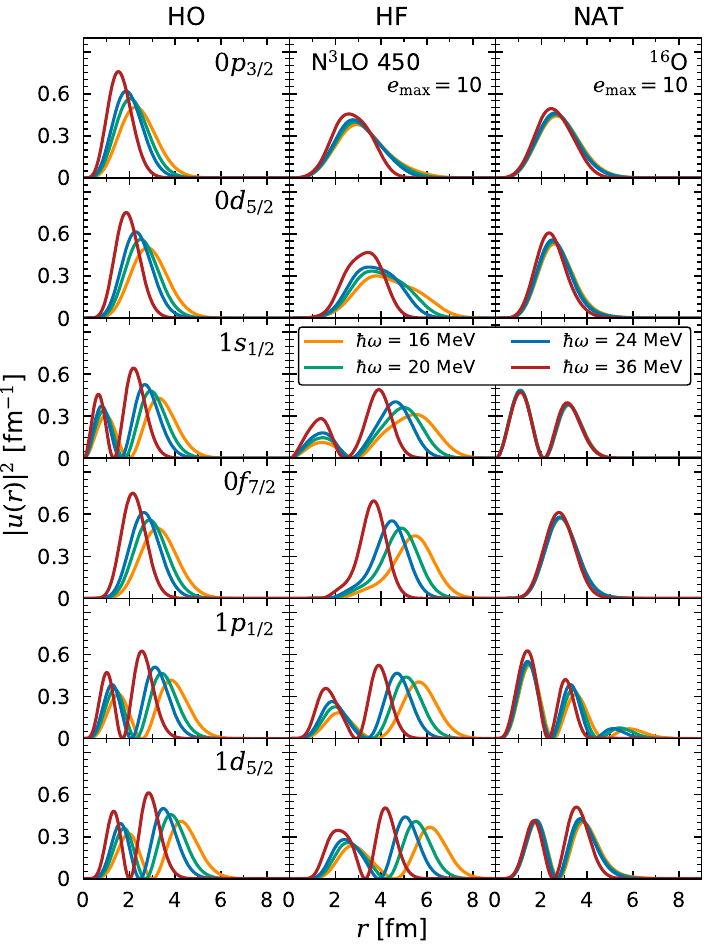}
\caption{Radial densities for harmonic-oscillator wave functions (HO), compared to Hartree-Fock (HF) and natural orbitals (NAT) for the nucleus $^{16}$O, shown for different frequencies $\hbar\omega$ of the underlying harmonic-oscillator basis. Figure taken from arXiv:2009.04701 with permission from the authors, see also Ref.~\cite{hoppe2021}.}
\label{fig:radialwaves}
\end{figure}

Instead of using a product state from Hartree-Fock theory, natural orbitals (i.e. the eigenstates of the one-body density matrix) are also quite attractive. This is because they are more localized than Hartree-Fock orbitals. As the nuclear force is short ranged, this facilitates including correlations beyond  Hartree-Fock. Tichai et al.~\cite{tichai2019} and Hoppe et al.~\cite{hoppe2021} showed that natural orbitals lead to ground-state energies that exhibit only a very mild dependence on the frequency of the underlying harmonic-oscillator basis. This can be seen by comparing squared radial single-particle wave functions. Figure~\ref{fig:radialwaves} compares these radial densities for harmonic-oscillator wave functions to  Hartree-Fock and natural orbitals for $^{16}$O for different frequencies of the underlying harmonic-oscillator basis. The mild dependence of the natural orbitals on the oscillator frequency is evident

Novario et al.~\cite{novario2020} found that using properly ordered natural orbitals (instead of Hartree-Fock orbitals) facilitate the inclusion of 3p--3h excitations in coupled-cluster theory. The large number of 3p--3h amplitudes, which scales as ${\cal O}(A^3{\cal M}^3)$, poses significant memory and computing demands. Interestingly, ordering the 3p--3h amplitudes, e.g. by their single-particle energy away from the Fermi surface, does not really help limiting the number of excitations. In contrast, following Ref.~\cite{tichai2019} one can compute perturbative corrections to the trivial density matrix from Hartree Fock (where all orbitals have occupations one or zero) and then choose the reference state such that the largest $A$ occupations are used. This reference state's energy expectation value is not as low as the Hartree-Fock energy but that is no concern. The 1p--1h excitations in coupled-cluster theory will correct this. The main point is that one can now order 3p--3h amplitudes $t_{ijk}^{abc}$ in descending order of the occupation-number product $(1-n_i)(1- n_j)(1- n_k) n_a n_b n_c$ (where $n_p$ is the occupation of the natural orbital labelled by $p$), and neglect all amplitudes where this product is smaller than some small number $\varepsilon$. In practice, this yields an ordering that is with respect to the Fermi surface. However, the calculations include all hole states, and one should possibly only use the product $n_a n_b n_c$  of occupation numbers in the particle space.   

\begin{figure}[htb]
\centering
\includegraphics[width=0.70\textwidth]{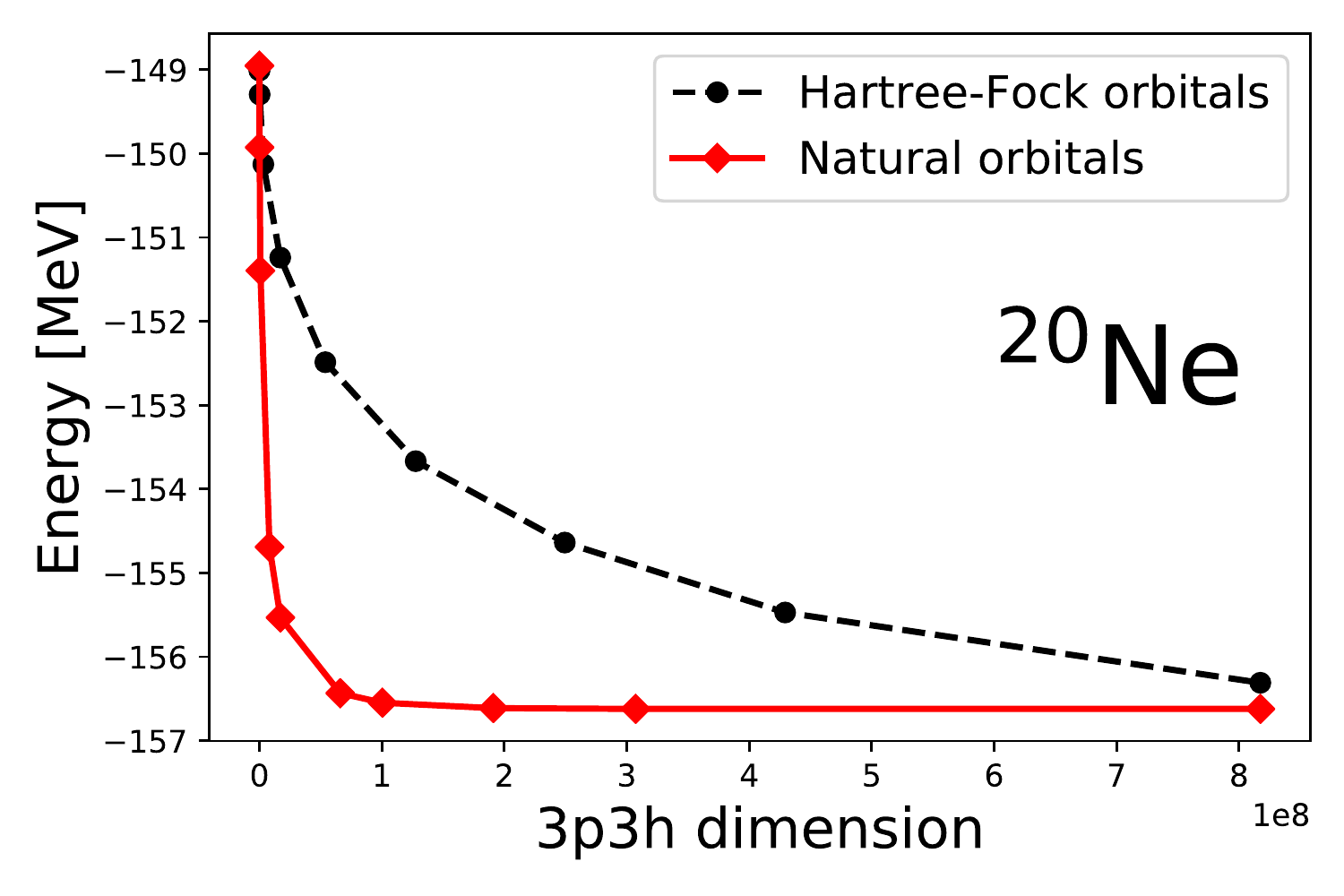}
\caption{Convergence of ground-state energy of $^{20}$Ne using coupled-cluster theory with approximate 3p--3h excitations as a function of the number of triples amplitudes (labeled as ``3p3h dimension''). Data taken from Ref.~\cite{novario2020}.}
\label{fig:natorbconv}
\end{figure}

Understanding of this fast convergence can also be based on Figure~\ref{fig:radialwaves}. The Hartree-Fock orbitals are clearly delocalized in position space, and for all but the largest oscillator frequency extend much beyond the size of the nucleus $^{16}$O. This is not unexpected as these wavefunctions are localized in energy. The natural orbitals, in contrast, are localized and have most of their support where it matters. Spatially localized orbitals, of course, combine well with the short-range nuclear interaction. Localized orbitals are useful to build short-range correlations into the wave function. This analysis is consistent with the arguments made in Sect.~\ref{sec:dividing}.

\subsection{Normal ordering with respect to the nontrivial vacuum state}
\label{sec:no}
The mean field computations yield a reference state. It is productive to consider this state as the nontrivial vacuum state of the $A$-body system. This invites one to normal-order the Hamiltonian with respect to this reference, i.e. rewriting the Hamiltonian such that all operators $\hat{a}^\dagger_i$ and $\hat{a}_a$ that annihilate the reference are to the right. 
This yields 
\begin{equation}
    H= E_{\rm ref} +H_{\rm no} \ ,
\end{equation}
where the energy $E_{\rm ref}$ of the reference state was defined in Eq.~(\ref{Eref}) and 
\begin{eqnarray}
\label{HNO}
    H_{\rm no}&\equiv& \sum_{pq} \langle q|H_{\rm no}|p\rangle \left\{\hat{a}^\dagger_q \hat{a}_p\right\} + {1\over 4} \sum_{pqrs} \langle pq|H_{\rm no}|rs\rangle \left\{\hat{a}^\dagger_p \hat{a}_q^\dagger \hat{a}_s \hat{a}_r\right\} \nonumber\\
    &&+{1\over 36} \sum_{pqrsuv}\langle pqu|H|rsv\rangle \left\{\hat{a}^\dagger_p \hat{a}_q^\dagger \hat{a}_u^\dagger\hat{a}_v\hat{a}_s \hat{a}_r \right\}
\end{eqnarray}
is the normal-ordered Hamiltonian. Here, the curly brackets $\{\cdots\}$ indicate normal ordering, one has $\langle\psi_0 |H_{\rm no}|\psi_0\rangle = 0$, and the matrix elements of the normal-ordered Hamiltonian are
\begin{eqnarray}
    \langle q|H_{\rm no}|p\rangle &=& \langle q|H|p\rangle +\sum_i \langle qi|H|pi\rangle + {1\over 2}\sum_{ij} \langle qij|H|pij\rangle \ ,\nonumber\\ 
    \langle pq|H_{\rm no}|rs\rangle &=& \langle pq|H|rs\rangle +  \sum_{i} \langle pqi|H|rsi\rangle \ .
\end{eqnarray}
Equation~(\ref{HNO}) shows that the two- and three-nucleon forces of the Hamiltonian $H$ contribute to the one- and  two-body forces of the normal-ordered Hamiltonian; they also contribute significantly to the reference energy $E_{\rm ref}$, see Eq.~(\ref{Eref}).  

The normal-ordering is important for the following two reasons. First, it severely limits the number of three-nucleon matrix elements required for the Hartree-Fock computation. Second, calculations of $^{4}$He~\cite{hagen2007a} and  of $^{16}$O~\cite{roth2012} showed that the normal-ordered three-body forces, i.e. the last line in Eq.~(\ref{HNO}) contributed very little to the ground-state energies of these nuclei~\footnote{``Very little'' means here that they contributed less than the uncertainty due to other approximations made in the computation.}. Based on those results, most practitioners neglect the  normal-ordered three-body forces in subsequent beyond-mean-field computations. This is the normal-ordered two-body approximation.

The reader may now wonder about the validity of the normal-ordered two-body approximation in deformed nuclei. There, the normal-ordered one-, two-, and three-body terms of $H_{\rm no}$ each break rotational invariance and only the sum of them yields a scalar under rotation. Thus, one cannot simply neglect the normal-ordered three-body forces in beyond-mean-field computations. Frosini et al.~\cite{Frosini:2021tuj} proposed a solution to this problem. They based the Hartree-Fock computation of an open-shell nucleus on a spherically symmetric one-body density matrix where valence orbitals are fractionally (and equally) occupied. Then, the normal-ordered Hamiltonian was computed using this density matrix, and the normal-ordered two-body approximation was made, i.e. the ``residual'' three-nucleon forces were omitted. Then, the resulting two-body Hamiltonian was transformed back into the original harmonic-oscillator basis. Such an Hamiltonian can then be used to compute a deformed reference state via Hartree Fock, and the normal-ordered Hamiltonian is computed with respect to that reference.  This two-body Hamiltonian can then be used for beyond-mean-field computations. The accuracy of this approach has been demonstrated in Ref.~\cite{Frosini:2021tuj}.
A similar solution for reference states that break particle number was presented in Ref.~\cite{ripoche2020}.

Let me end this Section with a philosophical note: Beyond-mean-field computations typically include two-particle--two-hole and three-particle--three-hole excitations. In some sense, they solve a generalized two- and three-body problem with respect to a non-trivial vacuum state. When viewed from this perspective, ab initio computations of $A$-body systems are conceptually not so different from few-body calculations that start from the true vacuum.

\section{Correlations}
\label{sec:ccm}
\subsection{Coupled-cluster method as an illustrative example}
There  are various beyond-mean-field methods that include correlations into the reference state; examples are coupled-cluster theory~\cite{hagen2014}, Gorkov-Green's function methods~\cite{dickhoff2004,soma2013}, the in-medium similarity renormalization group (IMSRG)~\cite{hergert2016}. In what follows, I will use the coupled-cluster method as an illustrative example. I believe that the arguments made below could also be formulated using other approaches. 

Let me briefly review the key points of coupled-cluster theory and remind the reader of the convention regarding indices as stated in the text after Eq.~(\ref{refket}). The ground state is written as
\begin{equation}
    |\psi\rangle = e^T|\psi_0\rangle \ , 
\end{equation}
with the reference state $|\psi_0\rangle$ from Eq.~(\ref{refket}) and the cluster operator
\begin{align}
\label{Tcluster}
        T &\equiv T_1 +T_2 +T_3 +\ldots \nonumber\\
        &=\sum_{ia} t_i^a \hat{a}^\dagger_a \hat{a}_i+ {1\over 4} \sum_{ijab}  t_{ij}^{ab} \hat{a}^\dagger_a \hat{a}_b^\dagger \hat{a}_j \hat{a}_i +{1\over 36} \sum_{ijkabc}t_{ijk}^{abc} \hat{a}^\dagger_a \hat{a}_b^\dagger \hat{a}_c^\dagger\hat{a}_k\hat{a}_j \hat{a}_i +\ldots
\end{align}
consisting of one-particle--one-hole (1p--1h), two-particle--two-hole (2p--2h),   three-particle--three-hole (3p--3h) excitations and so on. For a nucleus of mass number $A$, the cluster operator ends at $T_A$. In practice one truncates $T=T_1+T_2$ or $T=T_1+T_2+T_3$ and speaks of the coupled-cluster singles doubles (CCSD) and  coupled-cluster singles doubles triples (CCSDT) approximations, respectively. The cluster amplitudes solve the coupled-cluster equations 
\begin{align}
\label{CCEq}
    \langle\psi_i^a|\overline{H}_{\rm no}|\psi_0\rangle &=0 \ , \\
    \langle\psi_{ij}^{ab}|\overline{H}_{\rm no}|\psi_0\rangle &=0 \ ,\\
    \langle\psi_{ijk}^{abc}|\overline{H}_{\rm no}|\psi_0\rangle &=0 \ ,\\
    &\vdots \nonumber\\ 
    \langle\psi_{i_1\cdots i_A}^{a_1\cdots a_A}|\overline{H}_{\rm no}|\psi_0\rangle &=0 \ .
\end{align}
Here, 
\begin{align}
    |\psi_i^a\rangle  &\equiv \hat{a}^\dagger_a \hat{a}_i |\psi_0\rangle \ , \\
    |\psi_{ij}^{ab}\rangle  &\equiv \hat{a}^\dagger_a \hat{a}_b^\dagger \hat{a}_j \hat{a}_i |\psi_0\rangle \ , 
\end{align}
are 1p--1h and 2p--2h excitations of the reference state (and so on), and 
\begin{equation}
\label{Hsim}
    \overline{H}_{\rm no} \equiv e^{-T} H_{\rm no} e^T 
\end{equation}
is the similarity transformed (normal-ordered) Hamiltonian.   One sees that the ``optimal'' coupled-cluster amplitudes are such that the similarity-transformed Hamiltonian~(\ref{Hsim}) decouples the reference state from its 1p--1h and 2p--2h excitations in the CCSD approximation. In the CCSDT approximation the similarity-transformed Hamiltonian also decouples the reference state from 3p--3h excitations. 

The cluster operators only annihilate hole states and create particle states;  they are not anti-Hermitian and the operator $e^T$ is not unitary. Thus, the similarity-transformed Hamiltonian~(\ref{Hsim}) is not Hermitian. This drawback of coupled-cluster theory is compensated by the fact that the similarity transformed Hamiltonian~(\ref{Hsim}) can be computed exactly because the Baker-Campbell-Hausdorff expansion 
\begin{align}
    e^{-T}H_{\rm no}e^T = H_{\rm no} + \left[H_{\rm no},T\right] + {1\over 2!}\left[\left[H_{\rm no},T\right],T\right] + {1\over 3!}\left[\left[\left[H_{\rm no},T\right],T\right],T\right] + \ldots 
\end{align}
terminates. Assuming that $H_{\rm no}$ is a two-body operator (as it is in the normal-ordered two-body approximation) the Baker-Campbell-Hausdorff expansion terminates at four-fold nested commutators for any truncation of the cluster operator $T$. 

In coupled-cluster theory, the ground-state energy is
\begin{equation}
    E_0= E_{\rm ref} + E_{\rm corr} \ , 
\end{equation}
where the correlation energy is 
\begin{equation}
    E_{\rm corr}\equiv \langle\psi_0|\overline{H}_{\rm no}|\psi_0\rangle \ .
\end{equation}

I continue to assume that $H_{\rm no}$ is a two-body operator. Then, only the $T_1$ and $T_2$ amplitudes enter the correlation energy, independent of the truncation level of $T$ beyond $T_2$. Thus, the cluster amplitudes beyond $T_2$ only enter the correlation energy only indirectly because they enter the 
coupled-cluster equations~(\ref{CCEq}) and thereby impact the $T_1$ and $T_2$ amplitudes. Thus, in the normal-ordered two-body approximation
\begin{equation}
\label{E2}
    E_{\rm corr} = \sum_{ia} t_i^a \langle i|H_{\rm no}|a\rangle + {1\over 4}\sum_{ijab}\left(t_{ij}^{ab}+t_i^a t_j^b-t_i^bt_j^a\right) \langle ij|H_{\rm no}|ab\rangle \ ,
\end{equation}
and this expression is independent of the truncation level of the cluster operator.

One can now discuss the numerical effort in solving the (non-linear) coupled-cluster equations~(\ref{CCEq}). In the CCSD and CCSDT approximations the computation of the required matrix elements of the similarity-transformed Hamiltonian~(\ref{Hsim}) scales as ${\cal O}(A^2{\cal M}^4)$ and ${\cal O}(A^3{\cal M}^6)$, respectively. This is expensive but affordable on leadership-class computing facilities. The largest model spaces that are currently employed consist of ${\cal M}\approx 4000$ single-particle states. In contrast, the Bogoliubov coupled-cluster method does not partition the space of single-particle orbitals~\cite{signoracci2015}. Then the above scaling estimates need to be replaced by ${\cal O}({\cal M}^6)$ and ${\cal O}({\cal M}^9)$ for the two-body and three-body approximations, respectively. Likewise, the IMSRG's similarity transformation is unitary and therefore includes all orbitals. It also scales as ${\cal O}({\cal M}^6)$ and ${\cal O}({\cal M}^9)$ when truncated at the two- and three-body cluster level, respectively~\cite{heinz2021}. As ${\cal M}\gg A$, these methods are  more expensive than the coupled-cluster method based on Hartree-Fock references. However, approximations that are more affordable have been proposed most recently~\cite{he2024}.

\subsection{Short-range correlations yield the bulk of nuclear binding} 
\label{sec:short}

The energy $E_{\rm ref}$ of the reference state is proportional to the mass number. One can extend that argument to the correlation energy $E_{\rm corr}$ and also demonstrate that 
short-range correlations yield the bulk of the nuclear binding energy. To see this, one considers Eq.~(\ref{E2}) in the localized version~(\ref{localized}) of the Hartree-Fock basis. 

The key is again that the nuclear interaction is short-ranged when compared to the average two-nucleon distance. Thus, long-range correlations, i.e. those  cluster amplitudes $t_i^a$ and $t_{ij}^{ab}$ where the distances between points $i,a$ and any pair of the points $i,j,a,b$, respectively, are larger than the range of the nuclear force, do not contribute to the correlation energy. Thus, only sufficiently short-range particle-hole amplitudes contribute to the binding energy. 

Let us now consider such short-range correlations. 
For a fixed hole state $\langle i|$ the matrix element $\langle i|H_{\rm no}|a\rangle$ is only non-zero for a small number of particle states $|a\rangle$ close to $|i\rangle$, i.e. $|\mathbf{r}_i-\mathbf{r}_a|$ must be smaller than the range of nuclear interaction. Thus, the contribution of the first sum in Eq.~(\ref{E2}) is proportional to $A$. For the two-body matrix element $\langle ij|H_{\rm no}|ab\rangle$ to be nonzero, $|j\rangle$ has to be in close distance to $|i\rangle$, i.e. $|\mathbf{r}_i-\mathbf{r}_j|$ must be smaller than the range of nuclear interaction, and so must be $|\mathbf{r}_a-\mathbf{r}_b|$ and $|\mathbf{r}_a-\mathbf{r}_i|$ and $|\mathbf{r}_b-\mathbf{r}_j|$. Thus, the last  sum also scale as $A$. 
This is the simple argument why short-range interactions yield ground-state energies that are extensive, i.e. proportional to the mass number $A$.  

Please note that these arguments can be extended to the case where one does not make the normal-ordered two-body approximation. Then, the triples amplitudes $t_{ijk}^{abc}$ also enter the correlation energy and the argument goes through because three-nucleon forces are short-ranged as well.

Interestingly, because the spin-isospin degeneracy is $g=4$ only up to four nucleons can be really close and one would thus conclude that, in the presence of three- and four-nucleon forces, up to 3p--3h and 4p--4h amplitudes, respectively, can yield significant contributions to the binding energy.  

The correlation energy~(\ref{E2}) clearly depends only the short-range parts of the 1p--1h and 2p--2h cluster amplitudes. How do higher-rank clusters (and long-range parts from $T_1$ and $T_2$ contribute to the binding energy? To understand this, one returns to the equations~(\ref{Tcluster}) that determine the cluster amplitudes. Inspection shows that cluster amplitudes from $T_{|k-1|}$ to $T_{|k+2|}$ enter in the equation $\langle\psi_{i_1\cdots i_k}^{a_1\cdots a_k}|\overline{H}_{\rm no}|\psi_0\rangle =0$ for a normal-ordered Hamiltonian with two-body forces. Thus, any truncation of the cluster amplitudes introduces inaccuracies. For example, the $T_1$ and $T_2$ amplitudes in the CCSD approximation differ from those in the CCSDT approximation. Thus, higher-order clusters contribute indirectly to the correlation energy~(\ref{E2}) through their  
impact onto $T_1$ and $T_2$. Higher-rank cluster amplitudes also include long-range correlations and thereby impact the short- and long-range correlations of $T_1$ and $T_2$. As an example please consider 4p--4h excitations where states in two pairs of hole states are close to each other while the pairs themselves are at a distance. These can contribute to the short-range part of 2p--2h excitations.  As another example please consider 4p--4h excitations where only one pair of hole states are close to each other. These can contribute to the long-range part of 2p--2h excitations.  

While long-range amplitudes do not contribute to the correlation energy, they do impact the similarity transformation~(\ref{Hsim}) of the Hamiltonian and of other observables. They thereby impact the energies of excited states and of other observables. This shows that one can imagine situations where the binding energy of a nucleus is accurately computed (because the short-range parts of $T_1$ and $T_2$ are rather accurate) while other long-range observables such as electric transition strengths are inaccurately computed (because the long-range parts of $T_1$ and $T_2$ are inaccurate).  

\begin{figure}[htb]
\centering
\includegraphics[width=0.60\textwidth]{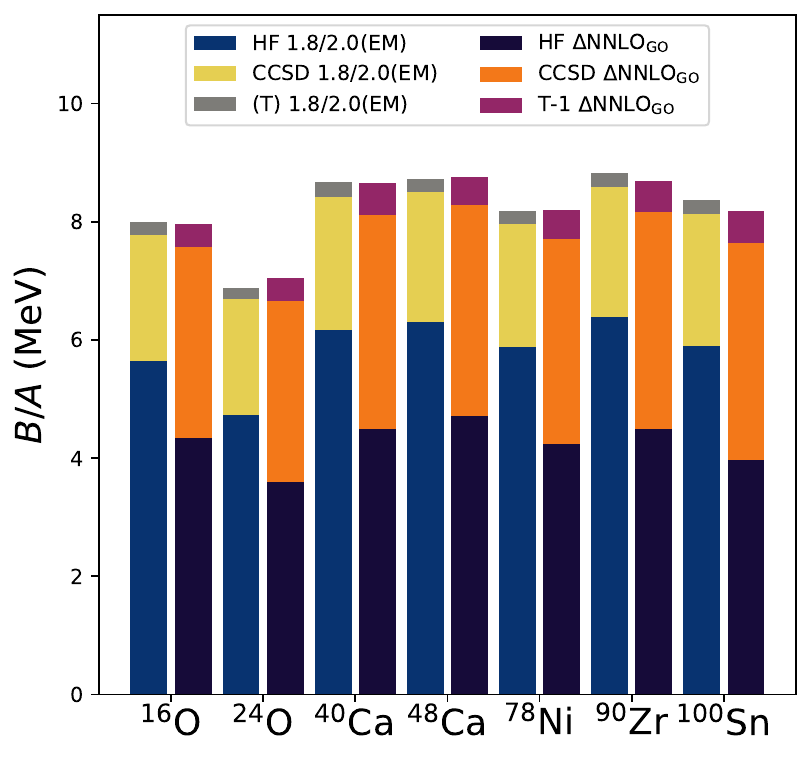}
\caption{Contributions to the binding energy of different nuclei from Hartree Fock (HF), coupled cluster with singles and doubles (CCSD) and triples (T or T-1). Results are shown for the chiral interaction 1.8/2.0(EM) in the bars on the left and for $\Delta$NNLO$_{\rm GO}$ as bars on the right for each nucleus. Data taken from Ref.~\cite{sun2022}.}
\label{fig:ccm-hierarchy}
\end{figure}

It is interesting to look at the contributions from singles ($T_1$), doubles $(T_2)$ and triples $(T_3)$ to the binding energy. Figure~\ref{fig:ccm-hierarchy} shows the results for a set of closed-shell nuclei using the interactions 1.8/2.0(EM) from Ref.~\cite{hebeler2011} and  $\Delta$NNLO$_{\rm GO}$ from Ref.~\cite{jiang2020}. 
Hartree Fock (denoted as HF) yields the largest contribution to the binding energy. The $T_3$ triples are included within the CCSD(T) approximation for the interaction 1.8/2.0(EM) and within the CCSDT-1 approximation for the interaction $\Delta$NNLO$_{\rm GO}$.
They are denoted as (T) and (T-1), respectively. In nuclear physics as in quantum chemistry~\cite{bartlett2007}, triples account for about 10\% of the correlation energy. One also sees that the 1.8/2.0(EM) is softer than the $\Delta$NNLO$_{\rm GO}$ because Hartree Fock yields a larger faction of the total binding energy. Indeed, the latter has a cutoff of 2~fm$^{-1}$ while the former has cutoffs 1.8~fm$^{-1}$ in the nucleon-nucleon and 2.0~fm$^{-1}$ in the three-nucleon sector.

\subsection{Renormalization of short-range correlations}

When encountering unknown or unpleasant short-range interactions, effective field theory states that the corresponding low-energy physics can be captured in a systematic way by adding contact potentials and derivatives thereof to the Hamiltonian~\cite{lepage1997}. The renormalization group reveals how the coupling strengths of these contacts change as the cutoff or resolution scale is changed~\cite{bogner2003,bogner2010}. 

Methods such as the coupled-cluster theory introduce 1p--1h, 2p--2h, 3p--3h, $\ldots$ correlations into the reference state, and the key contributions to the binding energy are short ranged.  What would happen if one ``integrated out'' some of these excitations, and how would one do that? Please imagine one started to remove short-range 3p--3h correlations that are above some momentum cutoff $\Lambda$. (To do this, one could for instance rewrite the $T_3$ cluster operator in hyperspherical coordinates and remove amplitudes at distances shorter than $\Lambda^{-1}$.) Doing so would, of course, change the ground-state energy in a many-body system, and lowering the cutoff to the Fermi momentum  would essentially remove all (short-ranged) triples amplitudes and one would change the correlation energy from CCSDT to CCSD. How could one possibly renormalize this removal of (short-ranged) 3p--3h correlations? Clearly, the renormalization can only affect three-body operators (and possibly operators of higher rank). Thus, proper renormalization would then require that, as one removes short-range physics in the 3p--3h sector, one compensates for that by adding a three-body contact operator and adjust its strength to one datum.  

\begin{figure}[htb]
\centering
\includegraphics[width=0.49\textwidth]{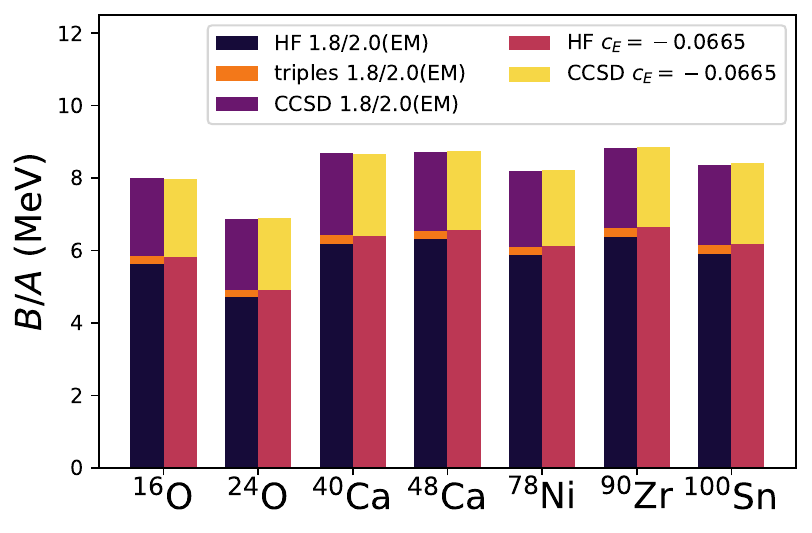}
\includegraphics[width=0.49\textwidth]{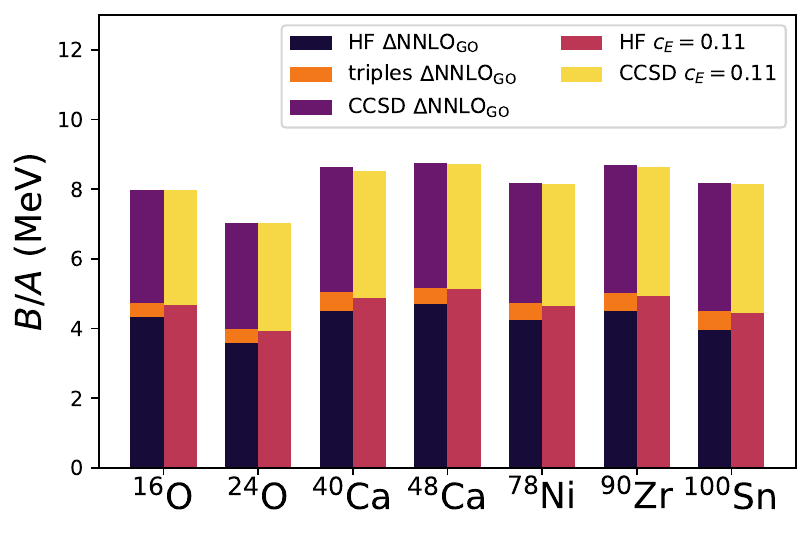}
\caption{Renormalization of 3p--3h amplitiudes in coupled-cluster calculations for the 1.8/2.0(EM) interaction from Ref.~\cite{hebeler2011} (left panel) and the $\Delta$NNLO$_{\rm GO}(398)$ potential from Ref.~\cite{jiang2020} (right panel). For each nucleus, the stacked three bars on the left show the contributions of Hartree-Fock (HF), 1p--1h and 2p--2h excitations (CCSD), and from 3p--3h excitations (triples) to the binding energy per nucleon $B/A$. The stacked two bars on the right show the contributions of Hartree-Fock (HF), 1p--1h and 2p--2h excitations (CCSD) of the renormalized interactions where the strength of the three-body contact $c_E$ has been adjusted to $^{16}$O. Data for the panels is taken from Ref.~\cite{sun2022}.}
\label{fig:CCSDrenorm}
\end{figure}

Sun et al.~\cite{sun2022} demonstrated that such a renormalization does indeed work. They performed coupled-cluster computations for $^{16}$O and included 3p--3h triples excitations. They also performed only CCSD calculations for that nucleus and adjusted the strength $c_E$ of the three-body contact such that those CCSD results with the renormalized $c_E$ match the previously computed result including triples excitations. They then computed other nuclei and compared triples results with those from CCSD using the three-body contact renormalized in $^{16}$O. The results obtained for two different interactions are shown in Fig.~\ref{fig:CCSDrenorm}. One sees that this renormalization works. Interestingly, the energy contribution from CCSD is virtually unchanged by the renormalization. Instead, the contribution of the Hartree-Fock energy almost completely accounts for the energy previously attributed to triples excitations.  These results also suggest that the bulk of 3p--3h excitations is short-ranged in nature. 

\subsection{Long-range correlations from symmetry projection}
\label{sub:long}
For all but doubly closed-shell nuclei, a mean-field state will break some of the continuous symmetries of the Hamiltonian. Then tools such as the coupled-cluster method can be used to include short-range correlations to account for the binding energy. However, such an approach (using a truncated cluster operator) cannot yield a spherical state when starting from a deformed reference. To see this, one assumes the reference state~(\ref{refket}) is deformed yet axially symmetric, and one takes the $z$ axis as the symmetry axis. Then, the state
\begin{align}
    |\Phi\rangle &\equiv \int\limits_0^{2\pi} {\rm d}\phi \int\limits_0^{\pi} {\rm d}\theta \sin\theta R(\phi,\theta) |\psi_0\rangle \nonumber\\
    &= \sum_{\alpha} R(\phi_\alpha,\theta_\alpha)|\psi_0\rangle 
\end{align}
is a state with zero angular momentum. Here, the rotation operator is $R(\phi,\theta)\equiv e^{-i\phi J_z} e^{-i\theta J_y}$ and in the second line it is assumed that one can replace the integration by a sum over angles $(\phi_\alpha,\theta_\alpha)$ that are equally distributed over the unit sphere. Using the Thouless theorem one can write 
\begin{align}
\label{sym1}
    |\Phi\rangle 
    &= \sum_{\alpha} w_\alpha e^{V_\alpha} |\psi_0\rangle
\end{align}
where the ``weights'' are $w_\alpha\equiv \langle \psi_0|R(\phi_\alpha,\theta_\alpha)|\psi_0\rangle$ and $V_\alpha$ is a one-body excitation operator, i.e. it has the same operator structure as the cluster operator $T_1$ in Eq.~(\ref{Tcluster}). One now writes the symmetry projected state as 
\begin{equation}
\label{sym2}
    |\Phi\rangle= e^{S_0+S_1+S_2+\ldots S_A}|\psi_0\rangle \ .
\end{equation}
Here, $S_r$, with $r=0,\ldots,A$ are $r$p--$r$h excitation operators (and $S_0$ is for normalization). Expanding the exponentials in Eqs.~(\ref{sym1}) and (\ref{sym2}) and equating the results yields
\begin{align}
    S_0&= \ln \sum_\alpha w_\alpha \ ,  \\
    S_1&= \langle V \rangle \ , \\
    S_2&= {1\over 2!}\left(\langle V^2\rangle - \langle V\rangle^2\right) \ , \\
    S_3&= {1\over 3!}\left(\langle V^3\rangle -3\langle V^2\rangle\langle V\rangle  +2\langle V\rangle^3\right) \ ,
\end{align}
and so on. Here,  $\langle f \rangle \equiv \sum_\alpha w_\alpha f_\alpha / \sum_\beta w_\beta$ for any operator $f_\alpha$. One sees that the excitation operator $S_r$ essentially is the $r$th cumulant of the weighted averages of the operators $V_\alpha$. Thus, all $S_r$,  $r=0,\ldots,A$ are expected to be non-zero and the symmetry restoration indeed involves up to $A$p--$A$h excitations, i.e., these correlations cannot be captured with a truncated cluster expansion. It is also clear that $V_\alpha$ induces long-range excitations, because $R(\phi_\alpha,\theta_\alpha)|\psi_0\rangle = w_\alpha e^{V_\alpha}|\psi_0\rangle$ is a rotated Slater determinant.

Two procedures are commonly used for symmetry projections~\cite{ringschuck,sheikh2021}. The first is based on group-theoretical techniques and the second exploits the generator coordinate method. For angular-momentum projections of axially symmetric states (with $J_z=0$) one can employ the energy functional
\begin{equation}
    E_J = \frac{\int\limits_0^{\pi} {\rm d}\theta \sin\theta d^J_{0,0}(\theta) {\cal H}(\theta) }{\int\limits_0^{\pi} {\rm d}\theta \sin\theta d^J_{0,0}(\theta) {\cal N}(\theta)} \ , 
\end{equation}
using the Wigner $d^J_{MK}(\theta)$ function~\cite{varshalovich1988}  and  the Hamiltonian and norm kernels
\begin{equation}
\label{kernelsHN}
\begin{split}
    {\cal H}(\theta) &\equiv \langle \tilde{\Psi}|H R(0,\theta)|\Psi\rangle \ , \\
    {\cal N}(\theta) &\equiv \langle \tilde{\Psi}|R(0,\theta)|\Psi\rangle \ .
\end{split}
\end{equation}
The evaluation of the norm kernels~(\ref{kernelsHN}) is only simple when $\langle \tilde{\Psi}|$ and $|\Psi\rangle$ are product states~\cite{robledo2020}. 
In general, the bra state $\langle \tilde{\Psi}|$ is not necessarily the adjoint of the ket $|\Psi\rangle$. The non-Hermitian coupled-cluster theory, for instance, parameterizes both states differently.  Qiu et al.~\cite{qiu2017} and Hagen et al.~\cite{hagen2022} used $|\Psi\rangle = e^T|\psi_0\rangle$ and $\langle\tilde\Psi| = \langle \psi_0| e^{-T}$ while Ref.~\cite{sun2024} employed the more sophisticated bra state 
$\langle\tilde\Psi| = \langle \psi_0| (1+\hat{\Lambda}) e^{-T}$ with $\hat{\Lambda}$ being a de-excitation operator.

An alternative (but closely related) approach to angular momentum projection follows the construction of effective Hamiltonians via the generator coordinate method~\cite{ringschuck}. This approach has the advantage that it can also be realized within an effective (field) theory. Let $|\Psi\rangle$ be a deformed, axially symmetric state and  
\begin{equation}
    |\Psi_\Omega\rangle \equiv R(\phi,\theta)|\Psi\rangle
\end{equation}
a rotation thereof. Here the shorthand  $\Omega\equiv(\phi,\theta)$ entered. One can compute the norm and Hamiltonian kernels
\begin{equation}
\label{HeffGCM}
\begin{split}
    {\cal N}_{\Omega'\Omega} &\equiv \langle\Psi_{\Omega'}|\Psi_\Omega\rangle \ , \\
    {\cal H}_{\Omega'\Omega} &\equiv \langle\Psi_{\Omega'}|H|\Psi_\Omega\rangle \ . 
\end{split}
\end{equation}
Solution of the generalized eigenvalue problem 
\begin{equation}
    \sum_\Omega {\cal H}_{\Omega'\Omega}\eta_\Omega = E \sum_\Omega {\cal N}_{\Omega'\Omega}\eta_\Omega
\end{equation}
then yields the projected energies and states; the corresponding value of $J$ can be inferred from degeneracy $2J+1$ of the computed states. 

As rotational motion is separated in scale from other excitations, one can also construct an effective (field) theory for these phenomena~\cite{papenbrock2011,papenbrock2020}. Instead of computing the nonlocal kernels~(\ref{HeffGCM}) for a given symmetry-breaking state and Hamiltonian, one posits a local norm kernel ${\cal N}_{\Omega'\Omega} =\delta(\Omega'-\Omega)$, and constructs a local effective Hamiltonian starting with the leading-order expression
\begin{equation}
    H_\Omega = E_0 - {1\over 2 C_0} \nabla_\Omega^2 \ .
\end{equation}
Here
\begin{equation}
\label{Heff}
    \nabla_\Omega \equiv \mathbf{e}_\theta \frac{\partial}{\partial\theta} + \mathbf{e}_\phi\frac{1}{\sin\theta} \frac{\partial}{\partial\phi}
\end{equation}
is the usual angular derivative using polar coordinates and unit vectors~\cite{varshalovich1988}, $E_0$ is the ground-state energy, and $C_0$ is the moment of inertia. Both parameters are low-energy constants that need to be adjusted to data or more microscopic calculations. One has $\nabla_\Omega^2=-\hat{J}^2$ where $\hat{J}$ is the angular momentum operator~\cite{varshalovich1988}. Thus, eigenfunctions of the effective Hamiltonian~(\ref{Heff}) are spherical harmonics $Y_{JM}(\Omega)$ and the corresponding energies are 
\begin{equation}
\label{rotspec}
    E_J = E_0 + \frac{J(J+1)}{2C_0} \ .
\end{equation}
For a symmetry-broken state the expectation value is
\begin{equation}
    \langle H_\Omega\rangle = E_0 +\frac{\langle \hat{J}^2\rangle}{2C_0} \ .
\end{equation}
Clearly, 
\begin{equation}
\label{eq:gain}
    \delta E \equiv E_0-\langle H_\Omega\rangle=-\langle \hat{J}^2\rangle/(2C_0)
\end{equation}
is the energy gain from angular momentum projection. 

The next point I want to make is that the restoration of rotational invariance does not yield a lot of energy. More precisely: the energy gain from symmetry projection is not size extensive and actually decreases with increasing mass number. Following Ref.~\cite{hagen2022}, the moment of inertia of the unprojected state scales as $C_0\sim A^{5/3}$ while $\langle \hat{J}^2\rangle  \sim A^{3/2}$ according to Ref.~\cite{bertsch2019}. Thus, the energy gain~(\ref{eq:gain}) from angular-momentum projection  decreases with increasing mass number. 

These arguments have been verified in ab initio computations where deformed reference states were projected onto states with good angular momentum~\cite{hagen2022}, and results from that work are compiled into Table~\ref{tab:gain}. One sees that the energy gained from angular momentum projection, $\delta E$ decreases with increasing mass numbers (both for the Hartree Fock and the coupled-cluster calculations). One also sees that the exciation energy of the $J^\pi=2^+$ state, $\Delta E \equiv E_2-E_0$ differs quite a bit between the Hartree-Fock and coupled-cluster results.  Here, the Hartree-Fock results are actually more accurate and closer to the benchmarks from the no-core shell model. This is because the symmetry-projected coupled-cluster calculations of Ref.~\cite{hagen2022} employ a very simple left (bra) state in the energy functional. A more sophisticated parameterization yields more accurate results~\cite{sun2024}.

\begin{table}[t]
\begin{tabular}{l|rrrrrrrr}     
Nucleus  & $E_{\rm HF}$& $E_{\rm CCD}$ & $\delta E_{\rm HF}$  & $\delta E_{\rm CCD}$ & $\Delta E_{\rm HF}$ & $\Delta E_{\rm CCD}$ &$\langle \hat{J}^2\rangle_{\rm HF}$&$\langle \hat{J}^2\rangle_{\rm CCD}$\\
          \hline
$^8$Be    & $-16.74$ & $ -50.24$   & $-7.4$ & $-3.3$ & 3.33 & 2.62&  $11.17$  &   $5.82$  \\
$^{20}$Ne & $-59.62$ & $-161.95$   & $-5.8$ & $-2.3$ & 1.26 & 1.19&  $21.26$  &   $12.09$ \\
$^{34}$Mg & $-90.21$ & $-264.34$   & $-3.2$ & $-1.5$ & 0.67 & 0.53&  $22.62$  &   $15.03$ \\
\end{tabular}
\caption{\label{tab:gain} Energies (all in MeV) and angular-momentum expectation values from unprojected Hartree-Fock (HF) and coupled-cluster (CCD) calculations, the energy gained from angular-momentum projection $\delta E$, and the excitation energy $\Delta E \equiv E_2-E_0$  of the $J^\pi=2^+$ state. Data taken from Ref.~\cite{hagen2022}. For $^8$Be and $^{20}$Ne, the no-core shell-model results are $\Delta E = 3.5$~\cite{caprio2015} and 1.64~MeV~\cite{dytrych2020}, and the experimental data are 3.03 and 1.63~MeV, respectively, and 0.66~MeV for $^{34}$Mg.} 
\end{table}

On first sight, it might be amazing that the angular-momentum projected Hartree-Fock results for $\Delta E$ in Table~\ref{tab:gain} are close to those from much more sophisticated (and expensive!) no-core shell-model computations. However, this is consistent with the arguments laid out below; rotational bands are long-wavelength physics and do not require high-resolution methods. The computation of the low-energy constant $E_0$ in the effective Hamiltonian~(\ref{Heff}) requires high resolution; the moment of inertia, $C_0$ requires much less resolution. For heavier nuclei, superfluidity becomes important, and this impacts the moment of inertia.  In those cases one can use symmetry-projected Hartree-Fock-Bogoliubov methods instead of Hartree-Fock~\cite{Frosini:2021sxj}. This is good news 
because symmetry projections of product states are relatively simple: Acting with a symmetry operation (such as a rotation) onto a product state simply yields another product state, and computing matrix elements between such states is straightforward and efficient~\cite{robledo2020}. Clearly, the approach via symmetry-breaking states is both conceptually and computationally simple.

\section{Excited states}
\label{sec:excited}
\subsection{Simple states from particle--hole excitations}
\label{sec:1p1h}

For an  understanding of excited states one best starts with heavy doubly magic nuclei such as $^{100}$Sn or $^{132}$Sn or $^{208}$Pb. In the Hartree-Fock basis one has the scaling relations~(\ref{scaleSPE}) for energy gaps between single-particle states and the average sizes of two-body matrix elements~(\ref{scaleME}). These relations are also valid for the corresponding one-body, two-body, and three-body matrix elements of the normal-ordered Hamiltonian. In heavy nuclei, these scaling relations clearly indicate a separation of scale between one-body, two body, and three-body matrix elements. This then suggests that there should be states that are dominated by 1p--1h excitations. Clearly, such states are conceptually and computationally simple.  

One writes excited states as 
\begin{equation}
    |R\rangle \equiv \hat{R}|\psi\rangle = \hat{R} e^T |\psi_0\rangle
\end{equation}
using the excitation operator
\begin{align}
\label{Rex}
        R &\equiv R_0 + R_1 +R_2 +R_3 +\ldots \nonumber\\
        &=R_0 + \sum_{ia} r_i^a \hat{a}^\dagger_a \hat{a}_i+ {1\over 4} \sum_{ijab}  r_{ij}^{ab} \hat{a}^\dagger_a \hat{a}_b^\dagger \hat{a}_j \hat{a}_i +{1\over 36} \sum_{ijkabc}r_{ijk}^{abc} \hat{a}^\dagger_a \hat{a}_b^\dagger \hat{a}_c^\dagger\hat{a}_k\hat{a}_j \hat{a}_i +\ldots \ .
\end{align}
The excitation operator is a spherical tensor of degree $J$ for the construction of excited states with spin $J$. Thus, one has $R_0=0$ for all excitations except when $J^\pi=0^+$. 

Let $E_{\rm ex}$ be the excitation energy of $|R\rangle$ with respect to the ground state. Then $|R\rangle$ fulfills the eigenvalue equation
\begin{equation}
\label{eigenR}
    \left[\overline{H}_{\rm no}, R\right]|\psi_0\rangle = E_{\rm ex} |\psi_0\rangle \ .
\end{equation}
Here, $\overline{H}_{\rm no}$ is the similarity transformed normal-ordered Hamiltonian  of Eq.~(\ref{Hsim}).

The interest is in an eigenstate with $J^\pi\ne 0^+$ and one assumes that this state is dominated by 1p--1h excitations. Thus, $R\approx R_1$ and the eigenvalue equation becomes
\begin{equation}
\label{eigenR1}
    \sum_b r_i^b \overline{H}_b^a -\sum_j r_j^a \overline{H}^j_i  + {1\over 2}\sum_{jb} \overline{H}_{ib}^{aj} r_j^b = E_{\rm ex} r_i^a
\end{equation}

As $T_1$ clusters only yield small contributions to $\overline{H}_{\rm no}$, one has (see Sect.~\ref{sec:dividing})
\begin{equation}
\begin{aligned}
\label{usefulEQ}
    \sum_b r_i^b \overline{H}_b^a -\sum_j r_j^a \overline{H}_i^j &\approx r_i^a \left(f_a^a - f_i^i \right) \\
    &\sim r_i^a  \Delta \varepsilon \ ,
    \end{aligned}
\end{equation}
where the shell gap $\Delta \varepsilon$ was introduced in Eq.~(\ref{shellgap}).
One seeks the low-energy solutions of the eigenvalue problem~(\ref{eigenR1}) and assumes that the dominant contributions come from excitations $r_i^a$ close to the Fermi surface. Then, 
\begin{equation}
    \sum_{kc} \overline{H}_{ic}^{ak} r_k^c \sim A^{-1/3} \times {\cal O}(r_i^a)
\end{equation}
because the sums over $k$ runs over the $A^{2/3}$ states in the shell just below the Fermi surface and $c$ is then fixed (because the aim is to create a 1p--1h excitation with the appropriate $J^\pi$); a typical matrix element is of size $A^{-1}$ (see Sect.~\ref{sec:dividing}). This estimate is conservative because of the assumption that all single-particle orbitals in the last occupied major shell (and the first unoccupied major shell) are degenerate. In reality there are fewer than ${\cal O}(A^{2/3})$ suitable states.  Thus, one would conclude from these arguments and from Eq.~(\ref{usefulEQ}) that the eigenvalue is of the size 
\begin{equation}
\label{Eex}
    E_{\rm ex} \sim \Delta \varepsilon \approx \frac{k_F^2}{2m} A^{-1/3} \ .
\end{equation}
This  explains the well-known fact  that low-lying excitations in sufficiently heavy nuclei can indeed be dominated by 1p--1h excitations, and that the corresponding energy is of the size of the shall gap. 

To obtain a more accurate excitation energy, one would need to include $R_2$ (and possibly $R_3$) into the excitation operator. Then, about twice the energy results from Eq.~(\ref{usefulEQ}), and this ultimately makes 2p--2h excitations a smaller correction to the dominant 1p--1h excitations. Similar arguments can then be made for particle-hole excitations of higher rank.

One also can extend these arguments towards low-lying $J^\pi=2^+$ states in closed-shell nuclei. In not too heavy nuclei (i.e. below the magic number 28), one needs 2p--2h excitations (because the parity changes at a shell closures). Thus, one sets $R=R_2$ in the eigenvalue problem~(\ref{eigenR}) and finds
\begin{equation}
\label{eigenR2}
    \sum_c r_{ij}^{ac} \overline{H}_c^b -\sum_k r_{ik}^{ab} \overline{H}^k_j  + {1\over 2}\sum_{kc} \overline{H}^{ak}_{ic} r_{kj}^{cb} + {1\over 2}\sum_{cd} \overline{H}^{ab}_{cd} r_{ij}^{cd} + {1\over 2}\sum_{kl} \overline{H}^{kl}_{ij} r_{kl}^{ab}= E_{\rm ex} r_{ij}^{ab} \ .
\end{equation}
Here terms to account for the proper antisymmetrization are suppressed because those are immaterial for the arguments that follow. For the first two sums, one argues as in the case of the 1p--1h excitations above (and again finds that those terms scale as the shell gap). For the last three double sums one notes that both the hole states and the particle states are confined to the last occupied or first unoccupied shell, respectively, and that there are  
${\cal O}(A^{1/3})$ single-$j$ orbitals in each of them. Thus, this is also the number to remove a spin-0 pair and to create a spin-2 pair.   Hamiltonian matrix elements  that induce this transition have a size of ${\cal O}(A^{-1})$, and again one arrives at a ${\cal O}(A^{-1/3})$ scaling for the contributions from these double sums.
Thus,  the 2p--2h excitations leading to $2^+$ states also have energies of the order of the shell gap, as in Eq.~(\ref{Eex}). 

The excited states discussed so far are dominated by few-particle--few-hole excitations  in the vicinity of the Fermi surface. Thus, they do not exhibit short-range correlations. This is also clear from the observation that these states differ in their orbital angular momentum and/or spin structure from the ground state. These are long-range structural changes with corresponding wave lengths that are of the order of the nuclear radius. This explains the well-known fact that computations of excitation energies (i.e. energy differences) require less resources than binding energies. Usually, excitation energies of such simple states converge much faster with increasing size of the model space than calculations of binding energies.  

\subsection{States in rotational bands}
Section~\ref{sub:long} already discussed how symmetry projections impact the ground-state energy. The symmetry projection then also yields excited states that belong to the ground-state rotational band. Using Eq.~(\ref{rotspec}) one sees that the excitation energies scale as ${\cal O}(A^{-5/3})$  estimated from the scaling of the moment of inertia. This scaling is indeed the lowest one can imagine: As the longest wave length scales as the radius, the corresponding momentum scales as $A^{1/3}$; taken together with a maximum collective mass that scales as $A$, the resulting kinetic energy scales as $A^{-5/3}$. Thus in sufficiently heavy nuclei, rotations are the lowest-energy excitations. While the understanding of these states is conceptually simple their computation is relatively expensive. In recent years, ab initio computations have been extended to include projection of angular momentum~\cite{sun2024,Frosini:2021sxj}.

\section{Results} 
\label{sec:impact}
What has been the impact of ab initio computations? The Green's function Monte Carlo~\cite{carlson1987,carlson1998,pieper2004,carlson2015} and no-core shell-model computations~\cite{navratil2000a,navratil2009,barrett2013,navratil2016} showed that properties of light nuclei (and their reactions) can accurately be computed from scratch by starting from two- and three-nucleon forces. In what follows I focus on the impact of scalable ab initio computations. The ability to link phenomena in medium-mass and heavy nuclei -- via chiral effective field theory -- to properties of nuclear forces rooted in quantum chromodynamics is exciting. In particular, this approach has a large predictive power. 
The advances of ab initio computations toward heavier nuclei allowed them to be confronted with many more experimental results and to help interpreting them, see, e.g., Refs.~\cite{garciaruiz2016,bagschi2020, koszorus2021,mougeot2021,ettenauer2022,koiwai2022,silwal2022,sommer2022,kaur2022,kondo2023,konig2023,chen2023,gray2023}.  This is another measure of their impact.
Ab initio computations have also led to genuinely new insights. In what follows, I will highlight a few examples.


\subsection{Neutron skins in $^{48}$Ca and $^{208}$Pb}
\label{subsec:skins}

\begin{figure}[htb]
\centering
\includegraphics[width=0.80\textwidth]{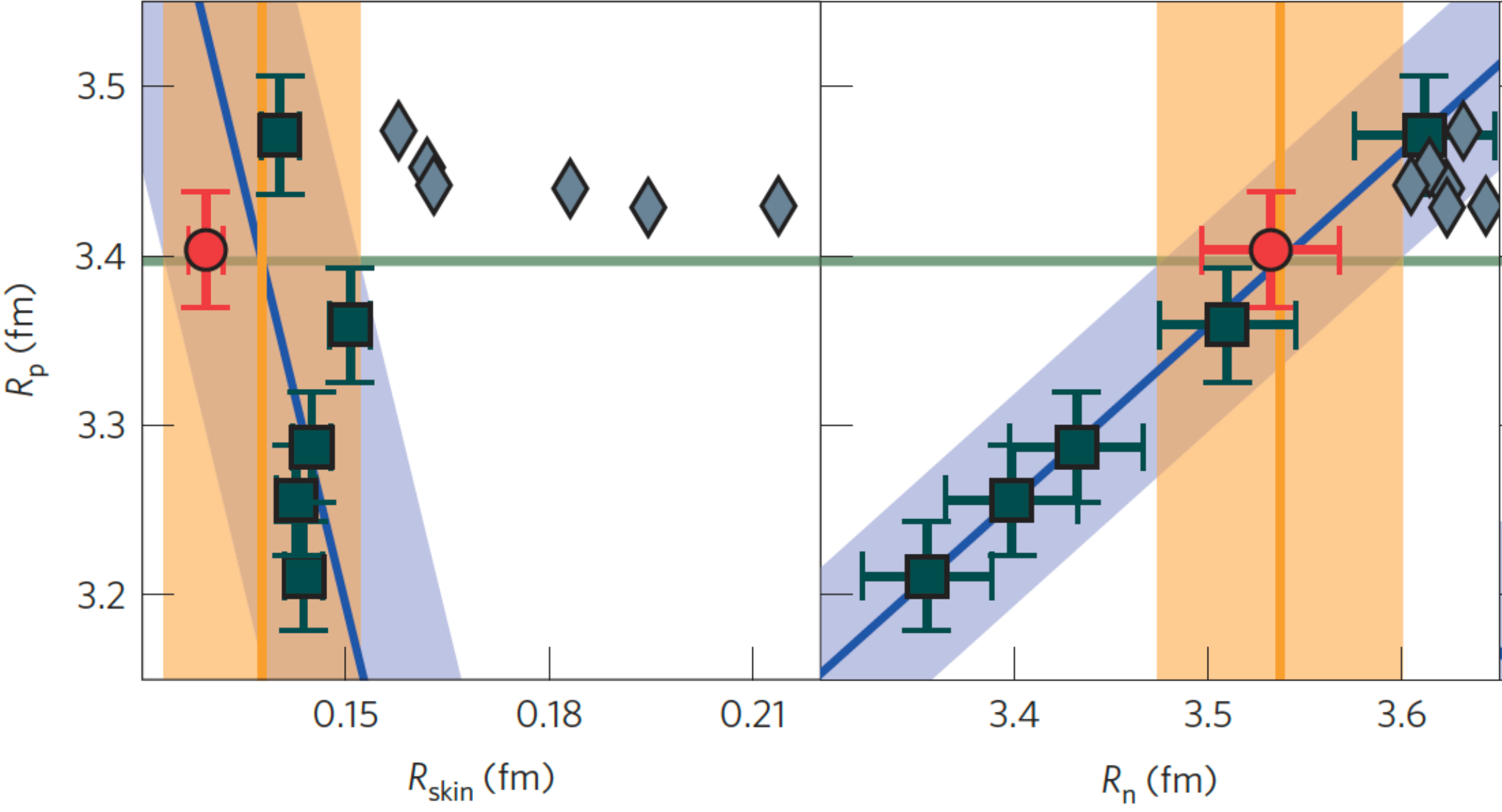}
\caption{Right panel: Correlation between the radii $R_p$ and $R_n$ of the proton and neutron distributions, respectively, in $^{48}$Ca, computed with interactions from chiral effective field theory (squares and circle with uncertainty estimates) and from nuclear energy density functionals (diamonds). Left panel: neutron-skin radius $R_{\rm skin}$ for the same interactions. The green horizontal line shows the precisely known value of $R_p$.  Extracted from arXiv:1509.07169 with permission from the authors, see Ref.~\cite{hagen2015} for details.}
\label{fig:48Ca-RNS}
\end{figure}

Most atomic nuclei contain more neutrons than protons, and this leads to the formation of a neutron skin at the nuclear surface. The neutron skin thickness $R_{\rm skin}$ is the difference between the radii of the neutron and proton distributions. This quantity is of interest, because a precise knowledge of  $R_{\rm skin}$ in nuclei allows one to constrain the neutron equation of state, thereby linking  properties of nuclei to those of neutron stars~\cite{brown2000,horowitz2001}. Charge radii of nuclei are well known from electron scattering or laser spectroscopy, but neutron distributions are harder to assess. Parity-violating electron scattering provides us with a model-independent way to measure the neutron skins because the exchanged $Z_0$~boson essentially only couples to the neutrons in a nucleus. In 2015, Hagen et al.~\cite{hagen2015} presented ab initio computations of the $^{48}$Ca neutron skin using interactions from chiral effective field theory. Their result $0.12~{\rm fm} \le R_{\rm skin} \le 0.15~{\rm fm}$ was later confirmed by the CREX experiment using parity-violating electron scattering~\cite{adhikari2022}. An interesting point of the theoretical prediction was that the neutron skin is a very robust quantity when computed with interaction from chiral effective field theory. This is shown in Fig.~\ref{fig:48Ca-RNS}. While the radii of the neutron and proton distributions depend strongly on the employed interaction, both quantities are strongly correlated and their difference (the neutron skin!) depends only weakly on the parameterization of the underlying Hamiltonian. This contrasts ab initio computations from calculations based on nuclear energy functionals.

More recently, ab initio computations~\cite{hu2022} also reported a result for the neutron skin in $^{208}$Pb. Those calculations quantified uncertainties and gave a 68\% credible interval of $0.14~{\rm fm} \le R_{\rm skin} \le 0.20~{\rm fm}$. This result is smaller than (and in mild tension with) the value extracted from parity-violating electron scattering by the PREX experiment~\cite{adhikari2021}. Interestingly, the theoretical uncertainty was smaller than the experimental one. The theorists found that constraints from $s$-wave scattering in the two-nucleon system prevented them from computing large neutron skins. The large neutron skin from the PREX experiment has puzzled theorists~\cite{reed2021,reinhard2021}, and it will be interesting to see the resolution of this tension.

\begin{figure}[htb]
\centering
\includegraphics[width=0.70\textwidth]{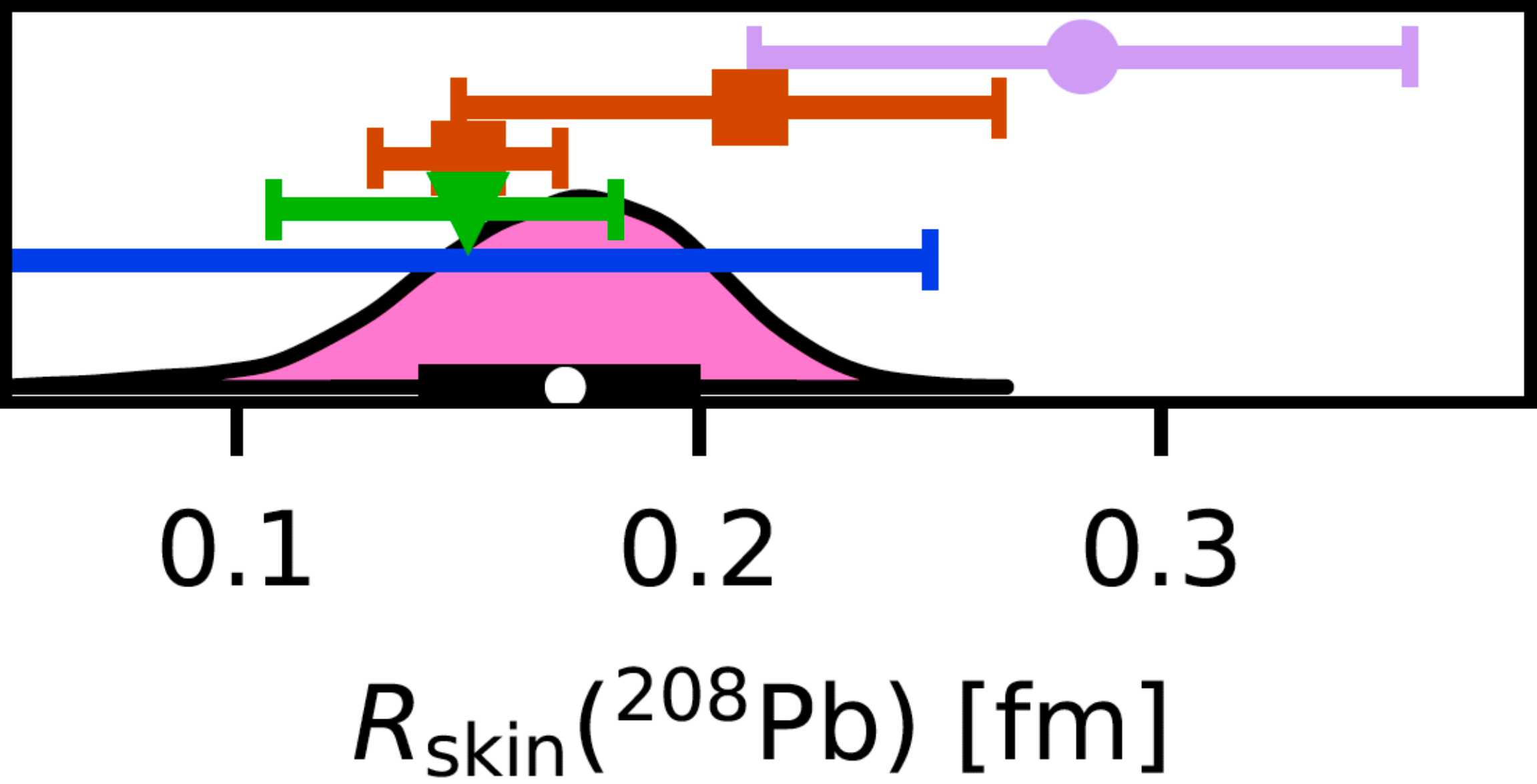}
\caption{Posterior predictive distribution for the neutron skin in $^{208}$Pb (with central point shown as a white dot and 68\% degree-of-belief interval as a horizontal black bar) compared to experimental extractions from PREX (purple), hadronic probes (red), electromagnetic probes
(green), and gravitational waves (blue). Extracted from arXiv:2112.01125 with permission from the authors, see Ref.~\cite{hu2022} for details.}
\label{fig:208Pb-RNS}
\end{figure}

\subsection{Quenching of $\beta$ decays}
\label{subsec:quench}
When computing $\beta$-decay rates in nuclei one typically finds that theory predicts faster rates than what is expected from the known $\beta$-decay of the free neutron. To achieve agreement with experiment, theory works if the Gamow-Teller transition strength is quenched by a factor $q^2\approx 0.75$, see, e.g., Refs.~\cite{chou1993,martinez1996}. This discrepancy has puzzled nuclear physicists for a long time. 

\begin{figure}[htb]
\centering
\includegraphics[width=0.60\textwidth]{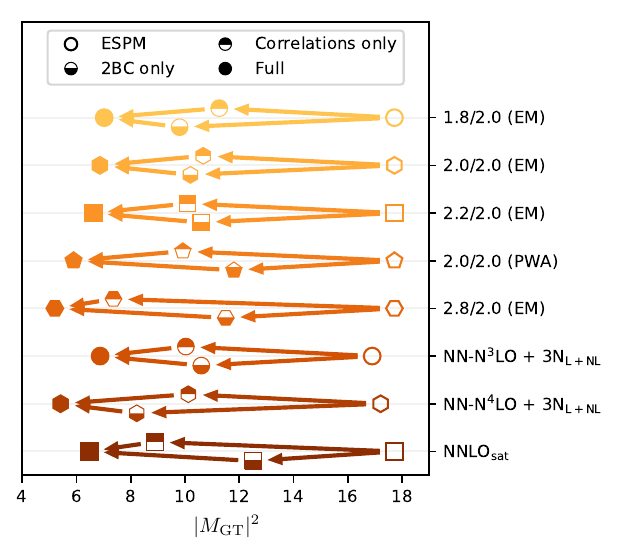}
\caption{Matrix element for Gamow-Teller transitions ($M_{\rm GT}$) in $^{100}$Sn computed with eight different interactions from chiral effective field theory. The extreme single-particle model (ESPM) yields a large matrix element, which is reduced by wavefunction correlations and two-body currents (2BC). Taken from arXiv:1903.00047 with permission from the authors, see Ref.~\cite{gysbers2019} for details.}
\label{fig:100SnQuench}
\end{figure}

When approached from chiral effective field theory, there is a subleading two-body current operator that serves as a correction to the Gamow-Teller transition operator. Its strength is set by the low-energy constants that appear in the leading three-nucleon forces. Gysbers et al.~\cite{gysbers2019} employed ab initio computations based on Hamiltonians and currents from chiral effective field theory. They computed $\beta$-decay rates in light and medium-mass nuclei up to $^{100}$Sn and found that the calculations agreed with data. Their analysis revealed that strong correlations in the nuclear wave function and two-body currents explain the apparent quenching of $\beta$ decays. 
This is shown in Fig.~\ref{fig:100SnQuench} for $^{100}$Sn.

\subsection{Matrix element for neutrinoless double-beta decay}
\label{subsec:doublebeta}
Neutrinoless double-$\beta$ decay is investigated worldwide in searches for physics beyond the Standard Model. If observed, it would tell us that the neutrino is its own antiparticle.  The lifetime of this decay is related to the unknown neutrino mass scale via a nuclear matrix element.~\cite{engel2017}. Thus matrix element must be computed and the challenge is that various nuclear models yield very different values for it. (A second challenge is that a contact operator of unknown strength is needed to properly renormalize this matrix element~\cite{cirigliano2018}.) Ab initio computations based on interactions from chiral effective field theory have started to address this problem~\cite{yao2020,belley2021,novario2021,belley2024}. The results from ab initio computations tend to be somewhat smaller than those from other models, and it will be interesting to see how much the unknown contact will contribute.  

\begin{figure}[htb]
\centering
\includegraphics[width=0.80\textwidth]{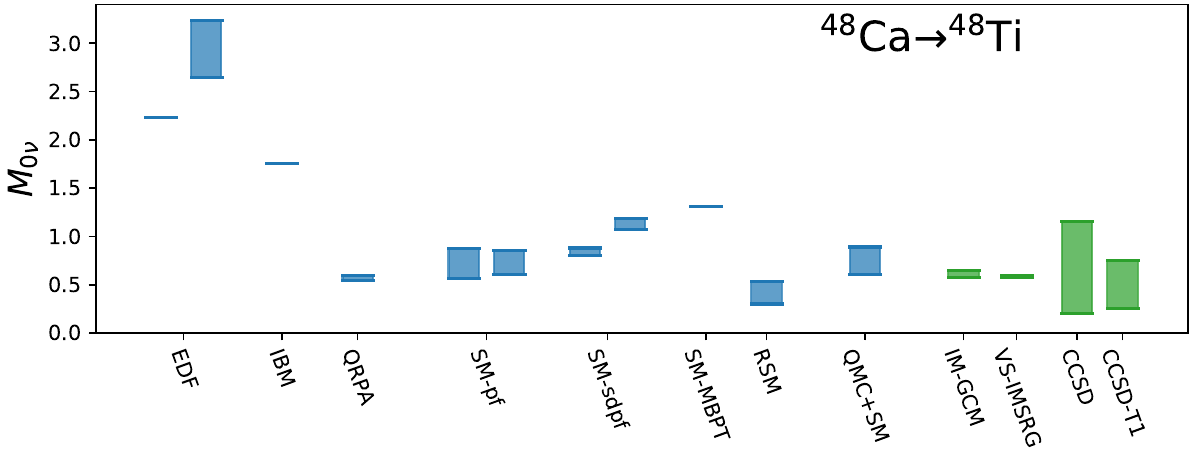}
\caption{Nuclear matrix element for neutrinoless double beta decay of $^{48}$Ca. Shown are the results of various models (in blue) compared with those from ab initio computations (green). Taken from arXiv:2207.01085 with permission from the authors, see Ref.~\cite{cirigliano2022} for details.}
\label{fig:0nubb}
\end{figure}

\section{Epilogue} 
\label{sec:summary}
Ab initio computations of atomic nuclei have advanced tremendously over the past two decades. These start from Hamiltonians that are based of effective field theories of quantum chromodynamics and use only controlled approximations. While the exact solution of such Hamiltonians might be exponentially expensive as the system size grows,  there are approximate solutions whose costs only grow  polynomially. These are a good match for Hamiltonians that themselves are only approximations of low-energy quantum chromodynamics.  The methods are  sufficiently precise to address interesting science questions. While these lectures were mainly focused on the coupled cluster method, several alternatives are available. The in-medium similarity renormalization group~\cite{tsukiyama2011}, Green's function~\cite{dickhoff2004}  and Gorkov methods~\cite{soma2013} are somewhat close in spirit, and much of what applies to coupled-cluster theory in terms of model spaces, reference states, particle-hole excitations, and correlations can be transferred to them. These methods exploit that many nuclei exhibit single reference states (or a family generated by a single reference via symmetry operations such as rotations). 

Nuclear lattice effective field theory~\cite{lee2009,laehde2019} is another method that scales well with mass number. It is based on a lattice (with periodic boundary conditions) in position space and uses quantum Monte Carlo techniques. A key of this approach consists of using an SU(4) spin-isospin symmetric Hamiltonian at leading order (to avoid or mitigate the fermion sign problem) and to apply other corrections perturbatively. This method is complementary to the other ab initio methods and has led to beautiful insights about the role of alpha-particles in nuclei and nuclear interactions~\cite{epelbaum2012,elhatisari2015,elhatisari2016,elhatisari2024}. 

In the past two decades, computing power has increased more than ten thousand fold. So it is no surprise that heavy deformed nuclei are being reached by ab initio methods~\cite{hu2024,door2024}. Beyond these advances in classical computing, quantum computing is emerging as an alternative technology~\cite{dumitrescu2018,farrell2024}. The outlook for ab initio computations of nuclei is bright.

\acknowledgments
I am grateful to all my collaborators and acknowledge stimulating discussions with many colleagues. I also thank the organizers and participants of the Varenna summer school for the wonderful atmosphere.  
This material is based upon work supported by the U.S.\ Department of Energy, Office of Science, Office of Nuclear Physics under award number DE-FG02-96ER40963. Oak Ridge National Laboratory is supported by the Office of Science of the Department of Energy under contract No. DE-AC05-00OR22725.

\end{document}